\documentclass[11pt,b4paper]{article}
\usepackage{bbm}
\usepackage{}
\usepackage{dsfont}
\usepackage{mathrsfs}
\usepackage{amssymb}
\usepackage{amsmath}
\usepackage{amsfonts}
\usepackage{amsthm}
\usepackage{epsfig}
\usepackage{enumerate}
\usepackage{graphicx}
\usepackage{epstopdf}
\usepackage{subfigure}
\usepackage{cite}

\newtheorem{lemma}{Lemma}

\newtheorem{remark}{Remark}

\newtheorem{assumption}{Assumption}

\long\def\comment#1{}

\title{Finite-Time Adaptive Fuzzy Tracking Control for Nonlinear State Constrained Pure-Feedback Systems\thanks {This work was supported in part by the National Natural Science Foundation
of China (61803122, 61873311), and the 111 Project (B16014).}
}

\author{Ju Wu\thanks{Research Institute of Intelligent Control and Systems, Harbin Institute of Technology, Harbin, 150080, China}, Tong Wang\thanks{Research Institute of Intelligent Control and Systems, Harbin Institute of Technology, Harbin, 150080, China}, Member, IEEE,
and Min Ma\thanks{Research Institute of Intelligent Control and Systems, Harbin Institute of Technology, Harbin, 150080, China}}

\date{}
\begin{document}

\maketitle \thispagestyle{plain}


\begin{abstract}
This paper investigates the finite-time adaptive fuzzy
tracking control problem for a class of pure-feedback system with
full-state constraints. With the help of Mean-Value Theorem,
the pure-feedback nonlinear system is transformed into strict-feedback case. By employing finite-time-stable like function
and state transformation for output tracking error, the output
tracking error converges to a predefined set in a fixed finite
interval. To tackle the problem of state constraints, integral
Barrier Lyapunov functions are utilized to guarantee that the
state variables remain within the prescribed constraints with
feasibility check. Fuzzy logic systems are utilized to approximate
the unknown nonlinear functions. In addition, all the signals
in the closed-loop system are guaranteed to be semi-global
ultimately uniformly bounded. Finally, two simulation examples
are given to show the effectiveness of the proposed control
strategy.


\end{abstract}



\section{Introduction}
In the past decades, the control of nonlinear systems have been paid considerable attention to. \cite{tong1} proposed fuzzy adaptive backstepping control for a class of nonlinear systems with uncertain unmodeled dynamics and disturbance. By introducing a modified Lyapunov function, \cite{zhang2000} designed an singularity-free controller based on NN for high-order strict-feedback nonlinear systems. \cite{wang2002} investigated adaptive neural network control for a class of SISO uncertain nonlinear systems in pure-feedback with backstepping technique.\cite{gepure2002} transformed nonaffine systems into affine systems with the help of mean theorem.\cite{tong2016} considered the case of immeasurable states, and proposed both fuzzy state feedback and observer-based output feedback control design. To overcome the so-called "explosion of complexity" problem induced by differentiating virtual control in traditional backstepping design, \cite{swaroop2000} first introduced dynamic surface control technique by designing low-pass filters.\cite{zhang2008} developed adaptive dynamic surface control for a class of pure-feedback nonlinear systems with unknown dead zone and perturbed uncertainties based on NN.\cite{meng2015}introduced a novel system transformation method that converts the nonaffine system into an affine system by combining state transformation and low-pass filter. Considering there exists a class of pure-feedback systems with nondifferentiable functions, \cite{liu2016} appropriately modeled the nonaffine functions without using mean value theorem. \cite{wangm2008} studied adaptive fuzzy control for uncertain SISO nonlinear time-delay systems in strict-feedback form, which was further generalized to nonaffine systems by \cite{wang2010}. To address input time delays, \cite{lih2017} and \cite{lid2019} employed Pade approximation techniques.

Constraints exist in almost all of physical systems. To avoid the performance degradation induced by violating constraints, effectively handling constraints in control design has been an important research topic practically and theoretically. \cite{blan1999} introduced invariant sets which laid the foundation for handling state and input constraints in linear systems. \cite{dehaan2005} designed extremum-seeking control for state-constrained nonlinear systems by using a barrier function. \cite{krstic2006} proposed a nonovershooting output tracking method for SISO strict-feedback nonlinear systems. \cite{tee2009} first designed barrier Lyapunov function (BLF) to prevent states from violating the constraints for nonlinear systems in strict feedback form. \cite{tee2011} used barrier Lyapunov function to solve partial state constraints problem. \cite{1meng2015} employed a error transformation method to tackle with time-varying output constraints for MIMO nonlinear systems. \cite{ren2010}, \cite{tang2016}, \cite{liuy2016} investigated adaptive NN control for uncertain nonlinear systems with full state constraints based on barrier Lyapunov functions, while less adjustable parameters are used by \cite{liuy2016}. By employing nonlinear mapping, \cite{zhang2017} transformed state constrained pure-feedback systems into novel pure-feedback systems without state constraints and designed adaptive NN controller without knowing control gain sign with the help of Nussbaum function. \cite{liuy2017} designed adaptive controller by the combination of BLF and Nussbaum function for state constrained nonlinear system with unknown control direction. Furthermore, \cite{tee2012} first introduced integral barrier Lyapunov function (iBLF) to simplify feasibility check. \cite{lid2016} proposed iBLF-based adaptive control for a class of affine nonlinear systems.

Most of the appropriately designed adaptive controllers make nonlinear systems satisfy ultimately uniform stability, thus driven by the need of manipulating systems to achieve prescribed performance in a finite interval, finite-time control has attracted remarkable attention. \cite{b2000} as a benchmark work of finite-time control studied the relationship between Lyapunov function and convergence time, which paved the way for solving many finite-time control problems of nonlinear systems. \cite{huang2005} obtained global finite-time stabilization for a class of uncertain nonlinear systems by adding a power integrator algorithm. \cite{wu2016} developed adaptive switching controller according to a novel Lyapunov-based switching rule for a class of nonlinear systems with multiple unknown control directions and global finite-time stabilization of the closed-loop systems was guaranteed. \cite{liuy2019} proposed a finite-time adaptive fuzzy tracking controller based on prescribed performance control and backstepping technique, which simplified the design process compared to previous works.

In this paper, we consider a class of perturbed state constrained pure-feedback nonlinear systems and construct finite-time adaptive fuzzy controller based on backstepping technique. By appropriately processing error transformation inspired by prescribed performance control with the help of finite-time-stable function, the output tracking error converges to preset arbitrarily small neighbor of the origin within a finite interval, and avoids violating predefined maximum overshoot. Integral barrier Lyapunov functions are employed to guarantee the states remain within preset constraints. Fuzzy logic systems are used to online approximate unknown system functions with tunable parameters. The main contributions of the proposed approach are that

(1)Up to now, few results before considered finite-time tracking control problem for state constrained pure-feedback nonlinear systems. Therefore a finite-time adaptive tracking controller is proposed for uncertain pure-feedback systems with state constraints and external perturbation. In the case of existing unknown control direction, the controller is redesigned to satisfy sufficient condition of stabilization proposed by \cite{ge2004}.

(2)The finite-time-stable function with the similar form to that introduced in \cite{b2000} is utilized to facilitate the error transformation. In the controller design, the function and its derivatives are employed as variables in fuzzy logic systems. Singularities are avoided by appropriately analyzing the relationship between the singularities of its derivatives and the parameters.

The rest of this paper is organized as follows.
Section II gives the problem formulation and preliminaries.
The finite-time adaptive fuzzy tracking design process is
given in Section III. Section IV presents Feasibility check. Two simulation examples are presented in
Section V to show the effectiveness of the proposed control
scheme. Finally, Section VI concludes this paper.
\section{Problem Statement and Preliminaries}
Considering the following pure-feedback system with full state constraints
\begin{equation}
{ \left\{ \begin{array}{l}
{{{\dot{x}}}_{i}}={{f}_{i}}({{{\bar{x}}}_{i}},{{x}_{i+1}})+{{d}_{i}}(t),i=1,2,\ldots ,n-1 \\
{{{\dot{x}}}_{n}}={{f}_{n}}({{{\bar{x}}}_{n}},u(t))+{{d}_{n}}(t) \\
y={{x}_{1}}
\end{array} \right.} \label{PFplant}
\end{equation}
where ${{\bar{x}}_{i}}={{[{{x}_{1}},{{x}_{2}},\cdots ,{{x}_{i}}]}^{T}}\in {{\mathbb{R}}^{i}},i=1,\ldots ,n$ and $x={{[{{x}_{1}},{{x}_{2}},\cdots ,{{x}_{n}}]}^{T}}\in {{\mathbb{R}}^{n}}$ are the state vectors of the system. $u(t)\in \mathbb{R}$, $y(t)\in \mathbb{R}$ and ${{d}_{i}}(t)\in \mathbb{R},i=1,\ldots ,n$ are input, output and external disturbances of the system respectively. And the following inequality holds:$|{{d}_{i}}(t)|\le {{D}_{i}},i=1,\ldots ,n$, where ${{D}_{i}}$ is unknown positive constant. ${{f}_{i}}({{\bar{x}}_{i}},{{x}_{i+1}}),i=1,\ldots ,n$ are unknown smooth nonlinear functions, and $y_d$ is the desired output signal.

The state variables are required to remain within prescribed constraints, i.e. $\vert x_i \vert \le {k_{ci}}, i=1,2, \ldots ,n$, where ${k_{ci}}$ is preset constant. In this paper, the control object is design an adaptive finite-time controller such that the output tracking error converges to a prescribed arbitrarily small neighbour of origin in a preset finite-time interval, the whole states remain within the predefined constraints and all the signals in the closed-loop system are bounded. To facilitate the controller design, we have the following basic knowledge.

\begin{lemma}
For a continuous function $\psi (x):\mathbb{R}^n \to \mathbb{R}$ which is defined on a compact ${{\Omega }_{x}}\in \mathbb{R}^n$, there exists a fuzzy logic system ${{W}^{T}}S(x)$ which can be used to approximate $\psi (x)$ with the technique including singleton, center average defuzzification and product inference, satisfying that
\begin{align}
\psi (x)={{W}^{T}}S(x)+&\varepsilon \\
\underset{x\in {{\Omega }_{x}}}{\mathop{\sup }}\,\left| \psi (x)-{{W}^{T}}S(x) \right|&\le {{\varepsilon }^{*}}
\end{align}
where $W={{[{{\omega }_{1}},{{\omega }_{2}},\ldots ,{{\omega }_{N}}]}^{T}}$ is the ideal weight vector, ${{\varepsilon }}$ is approximation error and ${{\varepsilon }^{*}}$ is unknown constant. $S(x)$ and $\xi (x)$ are basic functions and Gaussian functions respectively, which can be expressed as
\begin{align}
S(x)&=\frac{{{[{{\xi }_{1}}(x),{{\xi }_{2}}(x),...,{{\xi }_{N}}(x)]}^{T}}}{\sum\limits_{j=1}^{N}{{{\xi }_{j}}(x)}}, \\
\psi (x)&=\exp \left( \frac{-{{(x-{{l}_{j}})}^{T}}(x-{{l}_{j}})}{\eta _{j}^{T}{{\eta }_{j}}} \right)
\end{align}
where ${{l}_{j}}={{[{{l}_{j1}},{{l}_{j2}},\ldots ,{{l}_{jn}}]}^{T}}$ is the center vector, ${{\eta }_{j}}={{[{{\eta }_{j1}},{{\eta }_{j2}},\ldots ,{{\eta }_{jn}}]}^{T}}$ is the width of Gaussian function, while $n$ and $N$ are the number of system input and rules of fuzzy logic systems respectively.
\end{lemma}
\begin{lemma}
\cite{b2000} There exists the following finite-time-stable function satisfying
\begin{align}
\frac{d\vartheta \left( t \right)}{dt}=-\tau {{\left[ \vartheta \left( t \right) \right]}^{\kappa }},t\in \left[ 0,+\infty  \right),\label{DF}
\end{align}
where $\tau >0,0<\kappa <1$. solve the (\ref{DF}), we have
\begin{equation}
{\vartheta(t){=}\left\{ \begin{array}{l}
{{\left( {{\left( \vartheta \left( 0 \right) \right)}^{1-\kappa }}-\tau \left( 1-\kappa  \right)t \right)}^{\frac{1}{1-\kappa }}},t\in \left[ 0,{{T}_{0}} \right) \\
0,t\in \left[ {{T}_{0}},+\infty  \right)
\end{array} \right.} \label{DFS}
\end{equation}
where ${{T}_{0}}=\frac{{{\left( \vartheta \left( 0 \right) \right)}^{1-\kappa }}}{\tau \left( 1-\kappa  \right)}$. It's easy to see that if $\vartheta \left( 0 \right)>0$, then $\forall t\in \left[ 0,{{T}_{0}} \right),\vartheta \left( t \right)>0,\dot{\vartheta }\left( t \right)<0$. From (\ref{DFS}), we have $\underset{t\to {{T}_{0}}}{\mathop{\lim }}\,\vartheta \left( t \right)\text{=}0,\forall t\ge {{T}_{0}},\vartheta \left( t \right)\text{=}0$.
\end{lemma}
\begin{remark}
Since $\dot{\vartheta }\left( t \right)=-\tau {{\left( {{\left( \vartheta \left( 0 \right) \right)}^{1-\kappa }}-\tau \left( 1-\kappa  \right)t \right)}^{\frac{\kappa }{1-\kappa }}},t\in \left[ 0,{{T}_{0}} \right)$, it's necessary that $0<\kappa <1$ to avoid the singularity of $\dot{\vartheta }\left( t \right),t\to {{T}_{0}}$. Similarly, the $ith,i=2,\ldots ,n$ differential of ${\vartheta}(t)$ can be written as
\begin{align}
{{\vartheta }^{\left( i \right)}}\left( t \right)={{\left( -\tau  \right)}^{i}}\prod\limits_{j=1}^{i-1}{\left( j\kappa -j+1 \right)}{{\left( {{\left( \vartheta \left( 0 \right) \right)}^{1-\kappa }}-\tau \left( 1-\kappa  \right)t \right)}^{\frac{1}{1-\kappa }-i}},t\in \left[ 0,{{T}_{0}} \right)
\end{align}
The ${{\vartheta }^{\left( i \right)}}$ is involved in the following controller design. To avoid the possible singularity of ${{\vartheta }^{\left( i \right)}}$ when $t\to {{T}_{0}}$, select $1>\kappa >\frac{i-1}{i}$.
\end{remark}

\begin{lemma}
\cite{ge2004} $V\left( \cdot  \right)$ and $\zeta \left( \cdot  \right)$ are smooth functions defined on $t\in [0,{{t}_{f}})$, and $\forall t\in [0,{{t}_{f}}),V\left( t \right)\ge 0$. $N\left( \zeta  \right)$ is Nussbaum-type even function. If the following inequality holds
\begin{align}
0\le V\left( t \right)\le {{c}_{0}}+{{e}^{-{{c}_{1}}t}}\int_{0}^{t}{g\left( x\left( \tau  \right) \right)N\left( \zeta  \right)\dot{\zeta }{{e}^{{{c}_{1}}\tau }}}d\tau +{{e}^{-{{c}_{1}}t}}\int_{0}^{t}{\dot{\zeta }{{e}^{{{c}_{1}}\tau }}}d\tau ,\forall t\in [0,{{t}_{f}})
\end{align}
where $c_{0}$ and $c_{1}>0$ are suitable constants, and $g\left( x\left( \tau  \right) \right)$ is a time-varying parameter, which takes values in the unknown closed intervals $I=[{{l}^{-}},{{l}^{+}}]$, with $0\notin I$. Then $V(t)$, $\zeta(t)$ and $\int_{0}^{t}{g\left( x\left( \tau  \right) \right)N\left( \zeta  \right)\dot{\zeta }}d\tau $ must be bounded on $t\in [0,{{t}_{f}})$.
\end{lemma}

Define the output tracking error $z_{1}=x_{1}-y_{d}$, to guarantee the output error converges to the predefined arbitrarily small neighbor of origin in the prescribed finite-time interval, make error transformation as follows
\begin{align}
{{z}_{1}}=\mu \left( t \right)\Psi \left( e(t) \right) \label{ETF}
\end{align}
where $e(t)$ is a transformed error, $\mu(t)$ is finite-time-stable function and $\Psi\left( e(t) \right)\in[-1,1]$ is a smooth strictly increasing function satisfying $\underset{e(t)\to -\infty }{\mathop{\lim }}\,\Psi \left( e(t) \right)=-1$ and $\underset{e(t)\to +\infty }{\mathop{\lim }}\,\Psi \left( e(t) \right)=1$. We select $\Psi\left( e(t) \right)$ as $\frac{2}{\pi }\arctan \left( e(t) \right)$ in this paper. Inspired by Lemma 2, we yield $\mu \left( t \right)$ as
\begin{equation}
{\mu(t)=\left\{ \begin{array}{l}
{{\mu }_{{{T}_{0}}}}+{{(\mu _{0}^{\lambda }-\lambda \tau t )}^{\frac{1}{\lambda }}},t\in \left[ 0,{{T}_{0}} \right) \\
{{\mu }_{{{T}_{0}}}},t\in \left[ {{T}_{0}},+\infty  \right)
\end{array} \right.} \label{FTPF}
\end{equation}
where ${{\mu }_{{{T}_{0}}}}>0,{\tau}>0,1>\lambda >0$ are designed constants. It's easy to see that $\mu \left( 0 \right)={{\mu }_{{{T}_{0}}}}+{{\mu }_{0}}$, and ${{T}_{0}}={\mu _{0}^{\lambda }}/{\lambda \tau }$. $\mu \left( t \right)$ has the following finite-time-stable features: $\underset{t\to {{T}_{0}}}{\mathop{\lim }}\,\mu \left( t \right)={{\mu }_{{{T}_{0}}}}$, $\forall t>{{T}_{0}},\mu \left( t \right)={{\mu }_{{{T}_{0}}}}$. Due to the finite-time featured of $\mu(t)$, it's obvious that the output tracking error satisfying $\left| {{z}_{1}} \right|\le {{\mu }_{{{T}_{0}}}}$, when $t\ge {{T}_{0}}$. Since ${{\mu}^{\left( i \right)}},i=1,\ldots,n$ are involved in the following controller design, thus to avoid the possible singularity of ${{\mu }^{\left( i \right)}}$ when $t\to {{T}_{0}}$, select $0<\lambda <\frac{1}{n}$, where $n$ is the order of the pure-feedback system.

By Mean Theorem, the system (\ref{PFplant}) can be rewritten as
\begin{equation}
{ \left\{ \begin{array}{l}
{{{\dot{x}}}_{i}}={{f}_{i}}\left( {{{\bar{x}}}_{i}},0 \right)+{{g}_{i{{\iota }_{i}}}}{{x}_{i\text{+}1}}\text{+}{{d}_{i}}(t),1\le i\le n-1 \\
{{{\dot{x}}}_{n}}={{f}_{n}}\left( {{{\bar{x}}}_{n}},0 \right)\text{+}{{g}_{n{{\iota }_{n}}}}u(t)+{{d}_{n}}(t) \\
y={{x}_{1}}
\end{array} \right.} \label{PFplant2}
\end{equation}
where ${{g}_{i{{\iota }_{i}}}}={\partial {{f}_{i}}({{{\bar{x}}}_{i}},{{x}_{i+1}})}/{\partial {{x}_{i+1}}}\;{{|}_{{{x}_{i+1}}={{x}_{{\iota }{i}}}}}$ and ${{x}_{\iota i}}={{\iota }_{i}}{{x}_{i+1}}$. ${{\iota }_{i}},i=1,\ldots ,n$ are unknown constants satisfying $0<{{\iota }_{i}}<1$. Some commonly found assumptions are given as
\begin{assumption}
For the pure-feedback system (\ref{PFplant}), ${{g}_{i}}={\partial {{f}_{i}}({{{\bar{x}}}_{i}},{{x}_{i+1}})}/{\partial {{x}_{i+1}}}\;,i=1,\ldots ,n$, satisfying $0<{{g}_{i0}}<{{g}_{i}}<{{g}_{i1}},i=1,\ldots ,n$, where $g_{i0}$ and $g_{i1}$ are unknown constants in the set ${{\Omega }_{x}}=\left\{ x\in {{\mathbb{R}}^{n}}:\left| {{x}_{i}} \right|<{{k}_{ci}},i=1,\ldots ,n \right\}$.
\end{assumption}
\begin{assumption}
The desired output signal $y_d$ and its $i$-th derivative $y_d^{(i)}(t),i=1,\ldots ,n$ are known, continuous and bounded.
\end{assumption}

The time derivative of output tracking error $z_1$ is
\begin{align}
{{\dot{z}}_{1}}=\dot{\mu }\left( t \right)\Psi \left( e \right)+\mu \left( t \right)\frac{\partial \Psi \left( e(t) \right)}{\partial e(t)}\dot{e}(t), \label{DZ1}
\end{align}
which can be rewritten as
\begin{align}
\dot{e}(t)=\Phi \left( t \right)+\varphi \left( t \right){{\dot{z}}_{1}} \label{DZ2}
\end{align}
where $\Phi \left( t \right)=-\frac{\dot{\mu }\left( t \right)\Psi \left( e(t) \right)}{\mu \left( t \right){\partial \Psi \left( e(t) \right)}/{\partial e(t)}\;}$ and $\varphi \left( t \right)=\frac{1}{\mu \left( t \right){\partial \Psi \left( e(t) \right)}/{\partial e}\;(t)}$. According to transformed system (\ref{PFplant2}), the time derivative of $z_1$ is
\begin{align}
{{\dot{z}}_{1}}={{f}_{1}}({{x}_{1}},0)+{{g}_{1{{\iota }_{1}}}}{{x}_{2}}\text{+}{{d}_{1}}(t)-{{\dot{y}}_{d}}, \label{DZ3}
\end{align}
substitute (\ref{DZ3}) into (\ref{DZ2}), we obtain
\begin{align}
\dot{e}(t)=\Phi \left( t \right)+\varphi \left( t \right)\left( {{f}_{1}}({{x}_{1}},0)+{{g}_{1{{\iota }_{1}}}}{{x}_{2}}\text{+}{{d}_{1}}(t)-{{{\dot{y}}}_{d}} \right).
\end{align}
The transformed system (\ref{PFplant2}) can be further written as
\begin{equation}
{ \left\{ \begin{array}{l}
\dot{e}(t)=\Phi \left( t \right)+\varphi \left( t \right)\left( {{f}_{1}}({{x}_{1}},0)+{{g}_{1{{\iota }_{1}}}}{{x}_{2}}+{{d}_{1}}(t)-{{{\dot{y}}}_{d}}\right) \\
{{{\dot{x}}}_{i}}={{f}_{i}}\left( {{{\bar{x}}}_{i}},0 \right)+{{g}_{i{{\iota }_{i}}}}{{x}_{i\text{+}1}}\text{+}{{d}_{i}}(t),1\le i\le n-1 \\
{{{\dot{x}}}_{n}}={{f}_{n}}\left( {{{\bar{x}}}_{n}},0 \right)\text{+}{{g}_{n{{\iota }_{n}}}}u(t)+{{d}_{n}}(t) \\
\end{array} \right.} \label{PFplant2-1}
\end{equation}
\section{Controller Design}
In this section, finite-time adaptive fuzzy control laws will be designed based on the backstepping technique:

\textbf{Step $1$ :} Define the Lyapunov function as ${{V}_{e1}}={1}/{2}\;e{{(t)}^{2}}$ whose time derivative is
\begin{align}
{{\dot{V}}_{e1}}=e(t)\Phi \left( t \right)+e(t)\varphi \left( t \right)\left( {{f}_{1}}({{x}_{1}},0)+{{g}_{1{{\iota }_{1}}}}\left( {{z}_{2}}+{{\alpha }_{1}} \right)\text{+}{{d}_{1}}(t)-{{{\dot{y}}}_{d}} \right),\label{dV1}
\end{align}
where ${{z}_{2}}={{x}_{2}}-{{\alpha }_{1}}$, and $\alpha_{1}$ is the virtual control. Since ${{f}_{1}}({{x}_{1}},0)$ is unknown smooth function, with FLSs in Lemma 1, we have
\begin{align}
{{f}_{1}}({{x}_{1}},0)=W_{1}^{T}{{S}_{1}}\left( {{Z}_{1}} \right)+{{\varepsilon }_{1}}, \label{FLS1}
\end{align}
where $W_{1}$ is the optimal weight vector, ${{\varepsilon }_{1}}$ is the approximation error satisfying $\left| {{\varepsilon }_{1}} \right|\le \varepsilon _{1}^{*}$ and ${{Z}_{1}}={{x}_{1}}\in \mathbb{R}$. Substitute (\ref{FLS1}) into (\ref{dV1}), we have
\begin{align}
{{\dot{V}}_{e1}}=e(t)\Phi \left( t \right)+e(t)\varphi \left( t \right)\left( W_{1}^{T}{{S}_{1}}\left( {{Z}_{1}} \right)+{{\varepsilon }_{1}}+{{g}_{1{{\iota }_{1}}}}\left( {{z}_{2}}+{{\alpha }_{1}} \right)\text{+}{{d}_{1}}(t)-{{{\dot{y}}}_{d}} \right).\label{dV12}
\end{align}

By Young's inequality and Cauchy's inequality, we have
\begin{align}
e(t)\varphi W_{1}^{T}{{S}_{1}}\left( {{Z}_{1}} \right)&\le \frac{{{g}_{10}}e{{(t)}^{2}}\varphi _{1}^{2}{{\left\| {{W}_{1}} \right\|}^{2}}S_{1}^{T}{{S}_{1}}}{2a_{1}^{2}}+\frac{a_{1}^{2}}{2{{g}_{10}}} \label{I11} \\
-e(t)\varphi {{\dot{y}}_{d}}&\le \frac{{{g}_{10}}e{{(t)}^{2}}{{\varphi }^{2}}{{\left( {{{\dot{y}}}_{d}} \right)}^{2}}}{2}+\frac{1}{2{{g}_{10}}} \label{I12} \\
e(t)\varphi {{g}_{1{{\iota }_{1}}}}{{z}_{2}}&\le \frac{{{g}_{20}}e{{(t)}^{2}}{{\varphi }^{2}}z_{2}^{2}}{2}+\frac{g_{11}^{2}}{2{{g}_{20}}} \label{I13}\\
e(t)\Phi &\le \frac{{{g}_{10}}e{{(t)}^{2}}{{\Phi }^{2}}}{2}+\frac{1}{2{{g}_{10}}} \label{I14}\\
e(t)\varphi \left( {{\varepsilon }_{1}}\text{+}{{d}_{1}} \right)&\le {{g}_{10}}e{{(t)}^{2}}{{\varphi }^{2}}+\frac{\varepsilon {{_{1}^{*}}^{2}}+D_{1}^{2}}{2{{g}_{10}}} \label{I15}
\end{align}
where $a_1$ is designed positive constant. Define Lyapunov function as follows
\begin{align}
{{V}_{1}}={{V}_{e1}}+\frac{{{g}_{10}}}{2{{\beta }_{1}}}\tilde{\theta }_{1}^{2} \label{NLF1}
\end{align}
where $\beta_1$ is designed positive constant, ${{\tilde{\theta }}_{1}}=\theta _{1}^{*}-{{\hat{\theta }}_{1}}$, $\theta _{1}^{*}={{\left\| {{W}_{1}} \right\|}^{2}}$ and ${{\hat{\theta }}_{1}}$ is the approximation of $\theta _{1}^{*}$. Combined with inequalities (\ref{I11})-(\ref{I15}), the time derivative of (\ref{NLF1}) can be written as
\begin{align}
{{{\dot{V}}}_{1}}=&e(t)\Phi \left( t \right)+e(t)\varphi \left( t \right)\left( W_{1}^{T}{{S}_{1}}\left( {{Z}_{1}} \right)+{{\varepsilon }_{1}}+{{g}_{1{{\iota }_{1}}}}\left( {{z}_{2}}+{{\alpha }_{1}} \right)\text{+}{{d}_{1}}(t)-{{{\dot{y}}}_{d}} \right)-\frac{{{g}_{10}}}{{{\beta }_{1}}}{{{\tilde{\theta }}}_{1}}{{{\hat{\theta }}}_{1}} \nonumber \\
\le&e(t)\varphi \left( t \right){{g}_{1{{\iota }_{1}}}}{{\alpha }_{1}}+\frac{{{g}_{10}}e{{(t)}^{2}}{{\varphi }^{2}}{{\left\| {{W}_{1}} \right\|}^{2}}S_{1}^{T}{{S}_{1}}}{2a_{1}^{2}}+\frac{{{g}_{10}}e{{(t)}^{2}}{{\varphi }^{2}}{{\left( {{{\dot{y}}}_{d}} \right)}^{2}}}{2}+\frac{{{g}_{10}}e{{(t)}^{2}}{{\Phi }^{2}}}{2} \nonumber \\
&+{{g}_{10}}e{{(t)}^{2}}{{\varphi }^{2}}+\frac{{{g}_{20}}e{{(t)}^{2}}{{\varphi }^{2}}z_{2}^{2}}{2}+\frac{g_{11}^{2}}{2{{g}_{20}}}+\frac{1}{{{g}_{10}}}+\frac{a_{1}^{2}}{2{{g}_{10}}}+\frac{\varepsilon {{_{1}^{*}}^{2}}+D_{1}^{2}}{2{{g}_{10}}}-\frac{{{g}_{10}}}{{{\beta }_{1}}}{{{\tilde{\theta }}}_{1}}{{{\hat{\theta }}}_{1}} \label{dNLF1}
\end{align}

Design the virtual control and the adaptation parameter as
\begin{align}
{{\alpha }_{1}}&=-\frac{{{K}_{1}}e(t)}{\varphi }-\frac{{{{\hat{\theta }}}_{1}}e(t)\varphi S_{1}^{T}{{S}_{1}}}{2a_{1}^{2}}-\frac{e(t)\varphi {{\phi }_{1}}}{2}-e(t)\varphi -\frac{e(t){{\Phi }^{2}}}{2\varphi } \label{VC1} \\
{{\dot{\hat{\theta }}}_{1}}&=\frac{{{\beta }_{1}}e{{(t)}^{2}}{{\varphi }^{2}}S_{1}^{T}{{S}_{1}}}{2a_{1}^{2}}-{{\beta }_{1}}{{\sigma }_{1}}{{\hat{\theta }}_{1}} \label{AP1}
\end{align}
where ${{\phi }_{1}}={{\left( {{{\dot{y}}}_{d}} \right)}^{2}}$, $K_1>0$ and $\sigma_1$ are designed constants. It's easy to see $e(t)\varphi {{\alpha }_{1}}{{g}_{1{{\iota }_{1}}}}\le e(t)\varphi {{\alpha }_{1}}{{g}_{10}}$. Substituting (\ref{VC1}) and (\ref{AP1}) into (\ref{dNLF1}) obtains
\begin{align}
{{{\dot{V}}}_{1}}\le& e(t)\varphi {{g}_{1{{\iota }_{1}}}}\left( -\frac{{{K}_{1}}e(t)}{\varphi }-\frac{{{{\hat{\theta }}}_{1}}e(t)\varphi S_{1}^{T}{{S}_{1}}}{2a_{1}^{2}}-\frac{e(t)\varphi {{\phi }_{1}}}{2}-e(t)\varphi -\frac{e(t){{\Phi }^{2}}}{2\varphi } \right) \nonumber \\
&+\frac{{{g}_{10}}e{{(t)}^{2}}{{\varphi }^{2}}{{\left\| {{W}_{1}} \right\|}^{2}}S_{1}^{T}{{S}_{1}}}{2a_{1}^{2}}+\frac{{{g}_{10}}e{{(t)}^{2}}{{\varphi }^{2}}{{\left( {{{\dot{y}}}_{d}} \right)}^{2}}}{2}+\frac{{{g}_{10}}e{{(t)}^{2}}{{\Phi }^{2}}}{2}+\frac{{{g}_{20}}e{{(t)}^{2}}{{\varphi }^{2}}z_{2}^{2}}{2} \nonumber \\
&+{{g}_{10}}e{{(t)}^{2}}{{\varphi }^{2}}+\frac{g_{11}^{2}}{2{{g}_{20}}}+\frac{1}{{{g}_{10}}}+\frac{a_{1}^{2}}{2{{g}_{10}}}+\frac{\varepsilon {{_{1}^{*}}^{2}}+D_{1}^{2}}{2{{g}_{10}}}-\frac{{{g}_{10}}}{{{\beta }_{1}}}{{{\tilde{\theta }}}_{1}}\left( \frac{{{\beta }_{1}}e{{(t)}^{2}}{{\varphi }^{2}}S_{1}^{T}{{S}_{1}}}{2a_{1}^{2}}-{{\beta }_{1}}{{\sigma }_{1}}{{{\hat{\theta }}}_{1}} \right) \nonumber \\
\le&-{{K}_{1}}{{g}_{10}}e{{(t)}^{2}}+\frac{{{g}_{20}}e{{(t)}^{2}}{{\varphi }^{2}}z_{2}^{2}}{2}-{{g}_{10}}{{\sigma }_{1}}\frac{\tilde{\theta }_{1}^{2}}{2}+{{\Gamma }_{1}} \label{dNLF12}
\end{align}
where ${{\Gamma }_{1}}=\frac{g_{11}^{2}}{2{{g}_{20}}}+\frac{1}{{{g}_{10}}}+\frac{a_{1}^{2}}{2{{g}_{10}}}+\frac{\varepsilon {{_{1}^{*}}^{2}}+D_{1}^{2}}{2{{g}_{10}}}+{{g}_{10}}{{\sigma }_{1}}\frac{\theta _{1}^{*2}}{2}$.

\textbf{Step $2$ :} The time derivative of $z_2$ is
\begin{align}
{{\dot{z}}_{2}}={{f}_{2}}({{\bar{x}}_{2}},0)+{{g}_{2{{\iota }_{2}}}}{{x}_{3}}-{{\dot{\alpha }}_{1}}+{{d}_{2}}(t) \label{dZ2}
\end{align}

To guarantee the state variable remains within the preset constraint, define the integral Barrier Lyapunov function as
\begin{align}
{{V}_{z2}}=\int_{0}^{{{z}_{2}}}{\frac{\sigma k_{c2}^{2}}{k_{c2}^{2}-{{\left( \sigma +{{\alpha }_{1}} \right)}^{2}}}}d\sigma
\end{align}
substitute (\ref{dZ2}) into the time derivative of $V_{z2}$, we obtain
\begin{align}
 {{{\dot{V}}}_{z2}}&=\frac{\partial {{V}_{z2}}}{\partial {{z}_{2}}}{{{\dot{z}}}_{2}}+\frac{\partial {{V}_{z2}}}{\partial {{\alpha }_{1}}}{{{\dot{\alpha }}}_{1}} \nonumber \\
&=\frac{k_{c2}^{2}{{z}_{2}}}{k_{c2}^{2}-x_{2}^{2}}\left( {{f}_{2}}({{{\bar{x}}}_{2}},0)+{{g}_{2{{\iota }_{2}}}}{{x}_{3}}-{{{\dot{\alpha }}}_{1}}\text{+}{{d}_{2}}(t) \right)+\frac{\partial {{V}_{z2}}}{\partial {{\alpha }_{1}}}{{{\dot{\alpha }}}_{1}} \nonumber\\
&={{k}_{z2}}\left( W_{2}^{T}{{S}_{2}}\left( {{Z}_{2}} \right)+{{\varepsilon }_{2}}+{{g}_{2{{\iota }_{2}}}}\left( {{z}_{3}}+{{\alpha }_{2}} \right)-{{{\dot{\alpha }}}_{1}}+{{d}_{2}}(t) \right)+\frac{\partial {{V}_{z2}}}{\partial {{\alpha }_{1}}}{{{\dot{\alpha }}}_{1}} \label{dLF2}
\end{align}
where ${{k}_{z2}}=\frac{k_{c2}^{2}{{z}_{2}}}{k_{c2}^{2}-x_{2}^{2}}$, ${{z}_{3}}={{x}_{3}}-{{\alpha }_{2}}$ and $\alpha_2$ is virtual control. In accordance with Lemma 1, ${{f}_{2}}({{{\bar{x}}}_{2}},0)=W_{2}^{T}{{S}_{2}}\left( {{Z}_{2}} \right)+{{\varepsilon }_{2}}$, where $W_{2}$ is the optimal weight vector and ${{\varepsilon }_{2}}$ is approximation error, satisfying $\left|{{\varepsilon }_{2}}\right|\le {{\varepsilon }_{2}^{*}}$. ${{Z}_{2}}={{[{{x}_{1}},{{x}_{2}}]}^{T}}\in {{\mathbb{R}}^{2}}$.

Considering part of (\ref{dLF2})
\begin{align}
\frac{\partial {{V}_{z2}}}{\partial {{\alpha }_{1}}}{{{\dot{\alpha }}}_{1}}&={{{\dot{\alpha }}}_{1}}{{z}_{2}}\left( \frac{k_{c2}^{2}}{k_{c2}^{2}-x_{2}^{2}}-\int_{0}^{1}{\frac{k_{c2}^{2}}{k_{c2}^{2}-\left( \tau {{z}_{2}}+{{\alpha }_{1}} \right)}d\tau } \right) \nonumber \\
&={{{\dot{\alpha }}}_{1}}{{z}_{2}}\left( \frac{k_{c2}^{2}}{k_{c2}^{2}-x_{2}^{2}}-\frac{{{k}_{c2}}}{2{{z}_{2}}}\ln \frac{\left( {{k}_{c2}}+{{z}_{2}}+{{\alpha }_{1}} \right)\left( {{k}_{c2}}-{{\alpha }_{1}} \right)}{\left( {{k}_{c2}}-{{z}_{2}}-{{\alpha }_{1}} \right)\left( {{k}_{c2}}+{{\alpha }_{1}} \right)} \right) \nonumber \\
&=\frac{k_{c2}^{2}{{{\dot{\alpha }}}_{1}}{{z}_{2}}}{k_{c2}^{2}-x_{2}^{2}}-{{{\dot{\alpha }}}_{1}}{{z}_{2}}{{\rho }_{1}} \label{pdLF2}
\end{align}
where ${{\rho }_{1}}=\frac{{{k}_{c2}}}{2{{z}_{2}}}\ln \frac{\left( {{k}_{c2}}+{{z}_{2}}+{{\alpha }_{1}} \right)\left( {{k}_{c2}}-{{\alpha }_{1}} \right)}{\left( {{k}_{c2}}-{{z}_{2}}-{{\alpha }_{1}} \right)\left( {{k}_{c2}}+{{\alpha }_{1}} \right)}$. Since $\underset{{{\text{z}}_{2}}\to 0}{\mathop{\lim }}\,{{\rho }_{1}}=\frac{k_{c2}^{2}}{k_{c2}^{2}-\alpha _{1}^{2}}$, ${{\rho }_{1}}$ is well-defined in the neighbor of $z_2=0$, in the set $\left| {{\alpha }_{1}} \right|<{{k}_{c2}}$. Substituting (\ref{pdLF2}) into (\ref{dLF2}) yields
\begin{align}
{{\dot{V}}_{z2}}={{k}_{z2}}\left( W_{2}^{T}{{S}_{2}}\left( {{Z}_{2}} \right)+{{\varepsilon }_{2}}+{{g}_{2{{\iota }_{2}}}}\left( {{z}_{3}}+{{\alpha }_{2}} \right)\text{+}{{d}_{2}}(t) \right)-{{\dot{\alpha }}_{1}}{{z}_{2}}{{\rho }_{1}}
\end{align}
where ${{\dot{\alpha }}_{1}}=\frac{\partial {{\alpha }_{1}}}{\partial {{x}_{1}}}\left( W_{1}^{T}{{S}_{1}}\left( {{Z}_{1}} \right)+{{\varepsilon }_{1}}+{{g}_{1{{\iota }_{1}}}}{{x}_{2}}+{{d}_{1}}(t) \right)+\frac{\partial {{\alpha }_{1}}}{\partial {{y}_{d}}}{{\dot{y}}_{d}}+\frac{\partial {{\alpha }_{1}}}{\partial {{{\dot{y}}}_{d}}}{{\ddot{y}}_{d}}+\frac{\partial {{\alpha }_{1}}}{\partial \mu }\dot{\mu }+\frac{\partial {{\alpha }_{1}}}{\partial \dot{\mu }}\ddot{\mu }+\frac{\partial {{\alpha }_{1}}}{\partial {{{\hat{\theta }}}_{1}}}{{\dot{\hat{\theta }}}_{1}}$. Define $\theta _{2}^{*}=\max \left\{ {{\left\| {{W}_{1}} \right\|}^{2}},{{\left\| {{W}_{2}} \right\|}^{2}} \right\}$, by Young's inequality and Cauchy's inequality, the following inequalities are obtained
\begin{align}
{{k}_{z2}}W_{2}^{T}{{S}_{2}}\left( {{Z}_{2}} \right)&\le \frac{{{g}_{20}}k_{z2}^{2}\theta _{2}^{*}S_{2}^{T}{{S}_{2}}}{2a_{2}^{2}}+\frac{a_{2}^{2}}{2{{g}_{20}}}  \label{IE21}\\
{{k}_{z2}}{{g}_{2{{\iota }_{2}}}}{{z}_{3}}&\le \frac{{{g}_{30}}k_{z2}^{2}z_{3}^{2}}{2}+\frac{g_{21}^{2}}{2{{g}_{30}}} \\
{{k}_{z2}}\left( {{\varepsilon }_{2}}+{{d}_{2}} \right)&\le {{g}_{20}}k_{z2}^{2}+\frac{\varepsilon {{_{2}^{*}}^{2}}+D_{2}^{2}}{2{{g}_{20}}} \\
-{{z}_{2}}{{\rho }_{1}}\frac{\partial {{\alpha }_{1}}}{\partial {{x}_{1}}}W_{1}^{T}{{S}_{1}}\left( {{Z}_{1}} \right)&\le \frac{{{g}_{20}}z_{2}^{2}\rho _{1}^{2}\theta _{2}^{*}{{\left\| \frac{\partial {{\alpha }_{1}}}{\partial {{x}_{1}}}{{S}_{1}}\left( {{Z}_{1}} \right) \right\|}^{2}}}{2a_{2}^{2}}+\frac{a_{2}^{2}}{2{{g}_{20}}} \\
-{{z}_{2}}{{\rho }_{1}}\frac{\partial {{\alpha }_{1}}}{\partial {{x}_{1}}}{{g}_{1{{\iota }_{1}}}}{{x}_{2}}&\le \frac{{{g}_{20}}z_{2}^{2}\rho _{1}^{2}{{\left( \frac{\partial {{\alpha }_{1}}}{\partial {{x}_{1}}}{{x}_{2}} \right)}^{2}}}{2}+\frac{g_{11}^{2}}{2{{g}_{20}}} \\
-{{z}_{2}}{{\rho }_{1}}\frac{\partial {{\alpha }_{1}}}{\partial {{x}_{1}}}\left( {{\varepsilon }_{1}}+{{d}_{1}} \right)&\le {{g}_{20}}z_{2}^{2}{{\left( {{\rho }_{1}}\frac{\partial {{\alpha }_{1}}}{\partial {{x}_{1}}} \right)}^{2}}+\frac{\varepsilon {{_{1}^{*}}^{2}}+D_{1}^{2}}{2{{g}_{20}}} \\
-{{z}_{2}}{{\rho }_{1}}\sum\limits_{j=0}^{1}{\frac{\partial {{\alpha }_{1}}}{\partial y_{d}^{\left( j \right)}}y_{d}^{\left( j+1 \right)}}&\le \frac{{{g}_{20}}z_{2}^{2}\rho _{1}^{2}}{2}{{\sum\limits_{j=0}^{1}{\left( \frac{\partial {{\alpha }_{1}}}{\partial y_{d}^{\left( j \right)}}y_{d}^{\left( j+1 \right)} \right)}}^{2}}+\frac{1}{{{g}_{20}}} \\
-{{z}_{2}}{{\rho }_{1}}\sum\limits_{j=0}^{1}{\frac{\partial {{\alpha }_{1}}}{\partial {{\mu }^{\left( j \right)}}}{{\mu }^{\left( j+1 \right)}}}&\le \frac{{{g}_{20}}z_{2}^{2}\rho _{1}^{2}}{2}{{\sum\limits_{j=0}^{1}{\left( \frac{\partial {{\alpha }_{1}}}{\partial {{\mu }^{\left( j \right)}}}{{\mu }^{\left( j+1 \right)}} \right)}}^{2}}+\frac{1}{{{g}_{20}}} \\
-{{z}_{2}}{{\rho }_{1}}\frac{\partial {{\alpha }_{1}}}{\partial {{{\hat{\theta }}}_{1}}}{{\dot{\hat{\theta }}}_{1}}&\le \frac{{{g}_{20}}z_{2}^{2}\rho _{1}^{2}{{\left( \frac{\partial {{\alpha }_{1}}}{\partial {{{\hat{\theta }}}_{1}}}{{{\dot{\hat{\theta }}}}_{1}} \right)}^{2}}}{2}+\frac{1}{2{{g}_{20}}} \label{IE22}
\end{align}
where $a_2>0$ is a designed constant.

Design the following Lyapunov function
\begin{align}
{{V}_{2}}={{V}_{1}}+{{V}_{z2}}+\frac{{{g}_{20}}}{2{{\beta }_{2}}}\tilde{\theta }_{2}^{2}, \label{NLF2}
\end{align}
where $\beta_2$ is designed positive constant, ${{\tilde{\theta }}_{2}}=\theta _{2}^{*}-{{\hat{\theta }}_{2}}$, and ${{\hat{\theta }}_{2}}$ is the approximation of $\theta _{2}^{*}$. The time derivative of (\ref{NLF2}) is
\begin{align}
{{\dot{V}}_{2}}={{\dot{V}}_{1}}+{{k}_{z2}}\left( W_{2}^{T}{{S}_{2}}\left( {{Z}_{2}} \right)+{{\varepsilon }_{2}}+{{g}_{2{{\iota }_{2}}}}\left( {{z}_{3}}+{{\alpha }_{2}} \right)\text{+}{{d}_{2}}(t) \right)-{{\dot{\alpha }}_{1}}{z_2}{{\rho }_{1}}-\frac{{{g}_{20}}}{{{\beta }_{2}}}{{\tilde{\theta }}_{2}}{{\dot{\hat{\theta }}}_{2}} \label{dNLF2}
\end{align}

Design the virtual control as
\begin{align}
{{\alpha }_{2}}=-{{K}_{2}}{{z}_{2}}-\frac{e{{(t)}^{2}}\varphi^{2}{{z}_{2}}}{2}\left( \frac{k_{c2}^{2}-x_{2}^{2}}{k_{c2}^{2}} \right)-{{\phi }_{2}}-\frac{{{{\hat{\theta }}}_{2}}{{H }_{2}}}{2a_{2}^{2}} \label{VC2}
\end{align}
where $\phi_2$ and $H_2$ are functions of the signals from the first two subsystems, which can be expressed as
\begin{align}
{{\phi }_{2}}=&{{k}_{z2}}+\frac{{{z}_{2}}\rho _{1}^{2}}{2}{{\left( \frac{\partial {{\alpha }_{1}}}{\partial {{x}_{1}}}{{x}_{2}} \right)}^{2}}\left( \frac{k_{c2}^{2}-x_{2}^{2}}{k_{c2}^{2}} \right)+\frac{{{z}_{2}}\rho _{1}^{2}}{2}{{\sum\limits_{j=0}^{1}{\left( \frac{\partial {{\alpha }_{1}}}{\partial {{\pi }^{\left( j \right)}}}{{\pi }^{\left( j+1 \right)}} \right)}}^{2}}\left( \frac{k_{c2}^{2}-x_{2}^{2}}{k_{c2}^{2}} \right)  \nonumber \\
&+\frac{{{z}_{2}}\rho _{1}^{2}}{2}{{\sum\limits_{j=0}^{1}{\left( \frac{\partial {{\alpha }_{1}}}{\partial y_{d}^{\left( j \right)}}y_{d}^{\left( j+1 \right)} \right)}}^{2}}\left( \frac{k_{c2}^{2}-x_{2}^{2}}{k_{c2}^{2}} \right)+{{z}_{2}}\rho _{1}^{2}{{\left( \frac{\partial {{\alpha }_{1}}}{\partial {{x}_{1}}} \right)}^{2}}\left( \frac{k_{c2}^{2}-x_{2}^{2}}{k_{c2}^{2}} \right) \nonumber \\
&+\frac{{{z}_{2}}\rho _{1}^{2}}{2}{{\left( \frac{\partial {{\alpha }_{1}}}{\partial {{{\hat{\theta }}}_{1}}}{{{\dot{\hat{\theta }}}}_{1}} \right)}^{2}}\left( \frac{k_{c2}^{2}-x_{2}^{2}}{k_{c2}^{2}} \right) \label{phi2} \\
{{H }_{2}}=&{{k}_{z2}}S_{2}^{T}{{S}_{2}}+{{z}_{2}}\rho _{1}^{2}{{\left\| \frac{\partial {{\alpha }_{1}}}{\partial {{x}_{1}}}{{S}_{1}}\left( {{Z}_{1}} \right) \right\|}^{2}}\left( \frac{k_{c2}^{2}-x_{2}^{2}}{k_{c2}^{2}} \right) \label{H2}
\end{align}

Design the adaptation law as
\begin{align}
{{\dot{\hat{\theta }}}_{2}}=\frac{{{\beta }_{2}}{{k}_{z2}}{{H }_{2}}}{2a_{2}^{2}}-{{\beta }_{2}}{{\sigma }_{2}}{{\hat{\theta }}_{2}} \label{AP2}
\end{align}
where ${{\sigma }_{2}}>0$ is designed constant. It's easy to see that ${{k}_{z2}}{{g}_{2{{\iota }_{2}}}}{{\alpha }_{2}}\le {{k}_{z2}}{{g}_{20}}{{\alpha }_{2}}$. Substituting (\ref{IE21})-(\ref{IE22}), (\ref{VC2}) and (\ref{AP2}) into (\ref{dNLF2}) yields
\begin{align}
{{{\dot{V}}}_{2}}\le& {{{\dot{V}}}_{1}}+\frac{{{g}_{30}}k_{z2}^{2}z_{3}^{2}}{2}+\frac{{{g}_{20}}\theta _{2}^{*}{{H}_{2}}{{k}_{z2}}}{2a_{2}^{2}}+{{g}_{20}}{{k}_{z2}}{{\phi }_{2}}+{{k}_{z2}}{{g}_{20}}{{\alpha }_{2}}+\frac{5}{2{{g}_{20}}}+\frac{\varepsilon {{_{1}^{*}}^{2}}+D_{1}^{2}}{2{{g}_{20}}}  \nonumber \\
&+\frac{g_{11}^{2}}{{{g}_{20}}}+\frac{a_{2}^{2}}{2{{g}_{20}}}+\frac{\varepsilon {{_{2}^{*}}^{2}}+D_{2}^{2}}{2{{g}_{20}}}+\frac{g_{21}^{2}}{2{{g}_{30}}}-{{g}_{20}}{{{\tilde{\theta }}}_{2}}\left( \frac{{{k}_{z2}}{{H}_{2}}}{2a_{2}^{2}}-{{\sigma }_{2}}{{{\hat{\theta }}}_{2}} \right) \nonumber \\
=&{{{\dot{V}}}_{1}}+\frac{{{g}_{30}}k_{z2}^{2}z_{3}^{2}}{2}-{{K}_{2}}{{k}_{z2}}{{z}_{2}}{{g}_{20}}-\frac{{{g}_{20}}e{{(t)}^{2}}{{\varphi }^{2}}z_{2}^{2}}{2}-{{g}_{20}}{{\sigma }_{2}}\frac{\tilde{\theta }_{2}^{2}}{2}+{{\Gamma }_{2}} \label{dNLF22}
\end{align}
where ${{\Gamma }_{2}}=\frac{5}{2{{g}_{20}}}+\frac{\varepsilon {{_{1}^{*}}^{2}}+D_{1}^{2}}{2{{g}_{20}}}+\frac{g_{11}^{2}}{{{g}_{20}}}+\frac{a_{2}^{2}}{2{{g}_{20}}}+\frac{\varepsilon {{_{2}^{*}}^{2}}+D_{2}^{2}}{2{{g}_{20}}}+\frac{g_{21}^{2}}{2{{g}_{30}}}+{{g}_{20}}{{\sigma }_{2}}\frac{\theta _{2}^{*2}}{2}$. Substituting (\ref{dNLF12}) into (\ref{dNLF22}) yields
\begin{align}
{{\dot{V}}_{2}}\le -{{K}_{1}}{{g}_{10}}e{{(t)}^{2}}-{{K}_{2}}{{k}_{z2}}{{z}_{2}}{{g}_{20}}+\frac{{{g}_{30}}k_{z2}^{2}z_{3}^{2}}{2}-{{g}_{20}}{{\sigma }_{2}}\frac{\tilde{\theta }_{2}^{2}}{2}-{{g}_{10}}{{\sigma }_{1}}\frac{\tilde{\theta }_{1}^{2}}{2}+{{\Gamma }_{1}}+{{\Gamma }_{2}}
\end{align}

\textbf{Step $i$ :}($i=3,\cdots,n-1$) The time derivative of $z_i$ is
\begin{align}
{{\dot{z}}_{i}}={{f}_{i}}({{\bar{x}}_{i}},0)+{{g}_{i{{\iota }_{i}}}}{{x}_{i+1}}-{{\dot{\alpha }}_{i-1}}\text{+}{{d}_{i}}(t)
\end{align}
where ${{z}_{i}}={{x}_{i}}-{{\alpha }_{i-1}}$ and $\alpha_{i-1}$ is virtual control. Define the integral Barrier Lyapunov function as
\begin{align}
{{V}_{zi}}=\int_{0}^{{{z}_{i}}}{\frac{\sigma k_{ci}^{2}}{k_{ci}^{2}-{{\left( \sigma +{{\alpha }_{i-1}} \right)}^{2}}}}d\sigma, \label{ZI}
\end{align}
substitute (\ref{ZI}) into the time derivative of $V_{zi}$, we obtain
\begin{align}
{{{\dot{V}}}_{zi}}=&\frac{\partial {{V}_{zi}}}{\partial {{z}_{i}}}{{{\dot{z}}}_{i}}+\frac{\partial {{V}_{zi}}}{\partial {{\alpha }_{i-1}}}{{{\dot{\alpha }}}_{i-1}} \nonumber \\
=&\frac{k_{ci}^{2}{{z}_{i}}}{k_{ci}^{2}-x_{i}^{2}}\left( {{f}_{i}}({{{\bar{x}}}_{i}},0)+{{g}_{i{{\iota }_{i}}}}{{x}_{i+1}}-{{{\dot{\alpha }}}_{i-1}}\text{+}{{d}_{i}}(t) \right)+\frac{\partial {{V}_{zi}}}{\partial {{\alpha }_{i-1}}}{{{\dot{\alpha }}}_{i-1}} \nonumber \\
=&{{k}_{zi}}\left( W_{i}^{T}{{S}_{i}}\left( {{Z}_{i}} \right)+{{\varepsilon }_{i}}+{{g}_{i{{\iota }_{i}}}}\left( {{z}_{i+1}}+{{\alpha }_{i}} \right)-{{{\dot{\alpha }}}_{i-1}}\text{+}{{d}_{i}}(t) \right)+\frac{\partial {{V}_{zi}}}{\partial {{\alpha }_{i-1}}}{{{\dot{\alpha }}}_{i-1}} \label{dLFI}
\end{align}
where ${{k}_{zi}}=\frac{k_{ci}^{2}{{z}_{i}}}{k_{ci}^{2}-x_{i}^{2}}$, ${{z}_{i+1}}={{x}_{i+1}}-{{\alpha }_{i}}$ and $\alpha_i$ is virtual control. In accordance with Lemma 1, ${{f}_{i}}({{\bar{x}}_{i}},0)=W_{i}^{T}{{S}_{i}}\left( {{Z}_{i}} \right)+{{\varepsilon }_{i}}$, where $W_{i}$ is the optimal weight vector and ${{\varepsilon }_{i}}$ is approximation error, satisfying $\left|{{\varepsilon }_{i}}\right|\le {{\varepsilon }_{i}^{*}}$. ${{Z}_{i}}={{[{{x}_{1}},{{x}_{2}},\ldots ,{{x}_{i}}]}^{T}}\in {{\mathbb{R}}^{i}}$.

Considering part of (\ref{dLFI})
\begin{align}
\frac{\partial {{V}_{zi}}}{\partial {{\alpha }_{i-1}}}{{{\dot{\alpha }}}_{i-1}}=&{{{\dot{\alpha }}}_{i-1}}{{z}_{i}}\left( \frac{k_{ci}^{2}}{k_{ci}^{2}-x_{i}^{2}}-\int_{0}^{1}{\frac{k_{ci}^{2}}{k_{ci}^{2}-\left( \tau {{z}_{i}}+{{\alpha }_{i-1}} \right)}d\tau } \right) \\
=&{{{\dot{\alpha }}}_{i-1}}{{z}_{i}}\left( \frac{k_{ci}^{2}}{k_{ci}^{2}-x_{i}^{2}}-\frac{{{k}_{ci}}}{2{{z}_{i}}}\ln \frac{\left( {{k}_{ci}}+{{z}_{i}}+{{\alpha }_{i-1}} \right)\left( {{k}_{ci}}-{{\alpha }_{i-1}} \right)}{\left( {{k}_{ci}}-{{z}_{i}}-{{\alpha }_{i-1}} \right)\left( {{k}_{ci}}+{{\alpha }_{i-1}} \right)} \right) \\
=&\frac{k_{ci}^{2}{{{\dot{\alpha }}}_{i-1}}{{z}_{i}}}{k_{ci}^{2}-x_{i}^{2}}-{{{\dot{\alpha }}}_{i-1}}{{z}_{i}}{{\rho }_{i-1}} \label{pdLFI}
\end{align}
where ${{\rho }_{i-1}}=\frac{{{k}_{ci}}}{2{{z}_{i}}}\ln \frac{\left( {{k}_{ci}}+{{z}_{i}}+{{\alpha }_{i-1}} \right)\left( {{k}_{ci}}-{{\alpha }_{i-1}} \right)}{\left( {{k}_{ci}}-{{z}_{i}}-{{\alpha }_{i-1}} \right)\left( {{k}_{ci}}+{{\alpha }_{i-1}} \right)}$. Since $\underset{{{\text{z}}_{i}}\to 0}{\mathop{\lim }}\,{{\rho }_{i-1}}\text{=}\frac{k_{ci}^{2}}{k_{ci}^{2}-\alpha _{i-1}^{2}}$, ${{\rho }_{i}}$ is well-defined in the neighbor of $z_i=0$, when $\left| {{\alpha }_{i-1}} \right|<{{k}_{ci}}$. Substituting (\ref{pdLFI}) into (\ref{dLFI}) yields
\begin{align}
{{\dot{V}}_{zi}}={{k}_{zi}}\left( W_{i}^{T}{{S}_{i}}\left( {{Z}_{i}} \right)+{{\varepsilon }_{i}}+{{g}_{i{{\iota }_{i}}}}\left( {{z}_{i+1}}+{{\alpha }_{i}} \right)+{{d}_{i}}(t) \right)-{{\dot{\alpha }}_{i-1}}{{z}_{i}}{{\rho }_{i-1}}
\end{align}
where ${{\dot{\alpha }}_{i-1}}=\sum\limits_{j=1}^{i-1}{\frac{\partial {{\alpha }_{i-1}}}{\partial {{x}_{j}}}\left( W{{_{j}^{*}}^{T}}{{S}_{j}}\left( {{Z}_{j}} \right)+{{\varepsilon }_{j}}+{{g}_{j{{\iota }_{j}}}}{{x}_{j+1}}+{{d}_{j}} \right)}+\sum\limits_{j=0}^{i-1}{\frac{\partial {{\alpha }_{i-1}}}{\partial y_{d}^{\left( j \right)}}y_{d}^{\left( j+1 \right)}}+\sum\limits_{j=0}^{i-1}{\frac{\partial {{\alpha }_{i-1}}}{\partial {{\mu }^{\left( j \right)}}}{{\mu }^{\left( j+1 \right)}}} \\ +\sum\limits_{j=1}^{i-1}{\frac{\partial {{\alpha }_{i-1}}}{\partial {{{\hat{\theta }}}_{j}}}{{{\dot{\hat{\theta }}}}_{j}}}$. Define $\theta _{i}^{*}=\max \left\{ {{\left\| {{W}_{1}} \right\|}^{2}},{{\left\| {{W}_{2}} \right\|}^{2}},\ldots ,{{\left\| {{W}_{i}} \right\|}^{2}} \right\}$, by Young's inequality and Cauchy's inequality, the following inequalities are obtained
\begin{align}
{{k}_{zi}}W_{i}^{T}{{S}_{i}}\left( {{Z}_{i}} \right)&\le \frac{{{g}_{i0}}k_{zi}^{2}\theta _{i}^{*}S_{i}^{T}{{S}_{i}}}{2a_{i}^{2}}+\frac{a_{i}^{2}}{2{{g}_{i0}}} \label{IEI1} \\
{{k}_{zi}}{{g}_{i{{\iota }_{i}}}}{{z}_{i+1}}&\le \frac{{{g}_{\left( i+1 \right)0}}k_{zi}^{2}z_{i+1}^{2}}{2}+\frac{g_{i1}^{2}}{2{{g}_{\left( i+1 \right)0}}}  \\
{{k}_{zi}}\left( {{\varepsilon }_{i}}+{{d}_{i}} \right)&\le {{g}_{i0}}k_{zi}^{2}+\frac{\varepsilon {{_{i}^{*}}^{2}}+D_{i}^{2}}{2{{g}_{i0}}} \\
-{{z}_{i}}{{\rho }_{i-1}}\sum\limits_{j=1}^{i-1}{\frac{\partial {{\alpha }_{i-1}}}{\partial {{x}_{j}}}{{g}_{j{{\iota }_{j}}}}{{x}_{j+1}}}&\le \frac{{{g}_{i0}}z_{i}^{2}\rho _{i-1}^{2}}{2}\sum\limits_{j=1}^{i-1}{{{\left\| \frac{\partial {{\alpha }_{i-1}}}{\partial {{x}_{j}}}{{x}_{j+1}} \right\|}^{2}}}+\frac{1}{2{{g}_{i0}}}\sum\limits_{j=1}^{i-1}{g_{j1}^{2}} \\
-{{z}_{i}}{{\rho }_{i-1}}\sum\limits_{j=1}^{i-1}{\frac{\partial {{\alpha }_{i-1}}}{\partial {{x}_{j}}}\left( {{\varepsilon }_{j}}+{{d}_{j}} \right)}&\le {{g}_{i0}}z_{i}^{2}\rho _{i-1}^{2}\sum\limits_{j=1}^{i-1}{{{\left( \frac{\partial {{\alpha }_{i-1}}}{\partial {{x}_{j}}} \right)}^{2}}}+\frac{1}{2{{g}_{i0}}}\sum\limits_{j=1}^{i-1}{\left( \varepsilon {{_{j}^{*}}^{2}}+D_{j}^{2} \right)} \\
-{{z}_{i}}{{\rho }_{i-1}}\sum\limits_{j=0}^{i-1}{\frac{\partial {{\alpha }_{i-1}}}{\partial y_{d}^{\left( j \right)}}y_{d}^{\left( j+1 \right)}}&\le \frac{{{g}_{i0}}z_{i}^{2}\rho _{i-1}^{2}}{2}\sum\limits_{j=1}^{i-1}{{{\left\| \frac{\partial {{\alpha }_{i-1}}}{\partial y_{d}^{\left( j \right)}}y_{d}^{\left( j+1 \right)} \right\|}^{2}}}+\frac{i}{{{g}_{i0}}} \\
-{{z}_{i}}{{\rho }_{i-1}}\sum\limits_{j=0}^{i-1}{\frac{\partial {{\alpha }_{i-1}}}{\partial {{\mu }^{\left( j \right)}}}{{\mu }^{\left( j+1 \right)}}}&\le \frac{{{g}_{i0}}z_{i}^{2}\rho _{i-1}^{2}}{2}\sum\limits_{j=1}^{i-1}{{{\left\| \frac{\partial {{\alpha }_{i-1}}}{\partial {{\mu }^{\left( j \right)}}}{{\mu }^{\left( j+1 \right)}} \right\|}^{2}}}+\frac{i}{2{{g}_{i0}}} \\
-{{z}_{i}}{{\rho }_{i-1}}\sum\limits_{j=1}^{i-1}{\frac{\partial {{\alpha }_{i-1}}}{\partial {{{\hat{\theta }}}_{j}}}{{{\dot{\hat{\theta }}}}_{j}}}&\le \frac{{{g}_{i0}}z_{i}^{2}\rho _{i-1}^{2}}{2}\sum\limits_{j=1}^{i-1}{{{\left\| \frac{\partial {{\alpha }_{i-1}}}{\partial {{{\hat{\theta }}}_{j}}}{{{\dot{\hat{\theta }}}}_{j}} \right\|}^{2}}}+\frac{i-1}{2{{g}_{i0}}} \label{IEI2}
\end{align}
where $a_i>0$ is a designed constant.

Design the following Lyapunov function
\begin{align}
{{V}_{i}}={{V}_{i-1}}+{{V}_{zi}}+\frac{{{g}_{i0}}}{2{{\beta }_{i}}}\tilde{\theta }_{i}^{2}, \label{NLFi}
\end{align}
where $\beta_i$ is designed positive constant, ${{\tilde{\theta }}_{i}}=\theta _{i}^{*}-{{\hat{\theta }}_{i}}$, and ${{\hat{\theta }}_{i}}$ is the approximation of $\theta _{i}^{*}$. The time derivative of (\ref{NLFi}) is
\begin{align}
{{\dot{V}}_{i}}={{\dot{V}}_{i-1}}+{{k}_{zi}}\left( W_{i}^{T}{{S}_{i}}\left( {{Z}_{i}} \right)+{{\varepsilon }_{i}}+{{g}_{i{{\iota }_{i}}}}\left( {{z}_{i+1}}+{{\alpha }_{i}} \right)\text{+}{{d}_{i}}(t) \right)-{{\dot{\alpha }}_{i-1}}{z_i}{{\rho }_{i-1}}-\frac{{{g}_{i0}}}{{{\beta }_{i}}}{{\tilde{\theta }}_{i}}{{\dot{\hat{\theta }}}_{i}} \label{dNLFi}
\end{align}

Design the virtual control as
\begin{align}
{{\alpha }_{i}}=-{{K}_{i}}{{z}_{i}}-\frac{k_{z\left( i-1 \right)}^{2}{{z}_{i}}}{2}\left( \frac{k_{ci}^{2}-x_{i}^{2}}{k_{ci}^{2}} \right)-{{\phi }_{i}}-\frac{{{{\hat{\theta }}}_{i}}{{H}_{i}}}{2a_{i}^{2}} \label{VCi}
\end{align}
where $\phi_i$ and $H_i$ are functions of the signals from the first $i$ subsystems, which can be expressed as
\begin{align}
{{\phi }_{i}}=&{{k}_{zi}}+\frac{{{z}_{i}}\rho _{i-1}^{2}}{2}\sum\limits_{j=1}^{i-1}{{{\left\| \frac{\partial {{\alpha }_{i-1}}}{\partial {{x}_{j}}}{{x}_{j+1}} \right\|}^{2}}\left( \frac{k_{ci}^{2}-x_{i}^{2}}{k_{ci}^{2}} \right)}+{{z}_{i}}\rho _{i-1}^{2}\sum\limits_{j=1}^{i-1}{{{\left( \frac{\partial {{\alpha }_{i-1}}}{\partial {{x}_{j}}} \right)}^{2}}}\left( \frac{k_{ci}^{2}-x_{i}^{2}}{k_{ci}^{2}} \right) \nonumber \\
&+\frac{{{z}_{i}}\rho _{i-1}^{2}}{2}\sum\limits_{j=1}^{i-1}{{{\left\| \frac{\partial {{\alpha }_{i-1}}}{\partial y_{d}^{\left( j \right)}}y_{d}^{\left( j+1 \right)} \right\|}^{2}}\left( \frac{k_{ci}^{2}-x_{i}^{2}}{k_{ci}^{2}} \right)+}\frac{{{z}_{i}}\rho _{i-1}^{2}}{2}\sum\limits_{j=1}^{i-1}{{{\left\| \frac{\partial {{\alpha }_{i-1}}}{\partial {{\mu }^{\left( j \right)}}}{{\mu }^{\left( j+1 \right)}} \right\|}^{2}}}\left( \frac{k_{ci}^{2}-x_{i}^{2}}{k_{ci}^{2}} \right) \nonumber \\
&+\frac{{{z}_{i}}\rho _{i-1}^{2}}{2}\sum\limits_{j=1}^{i-1}{{{\left\| \frac{\partial {{\alpha }_{i-1}}}{\partial {{{\hat{\theta }}}_{j}}}{{{\dot{\hat{\theta }}}}_{j}} \right\|}^{2}}}\left( \frac{k_{ci}^{2}-x_{i}^{2}}{k_{ci}^{2}} \right) \\
{{H}_{i}}=&{{k}_{zi}}S_{i}^{T}{{S}_{i}}+{{z}_{i}}\rho _{i-1}^{2}\sum\limits_{j=1}^{i-1}{{{\left\| \frac{\partial {{\alpha }_{i-1}}}{\partial {{x}_{j}}}{{S}_{j}}\left( {{Z}_{j}} \right) \right\|}^{2}}}\left( \frac{k_{ci}^{2}-x_{i}^{2}}{k_{ci}^{2}} \right)
\end{align}

Design the adaptation law as
\begin{align}
{{\dot{\hat{\theta }}}_{i}}=\frac{{{\beta }_{i}}{{k}_{zi}}{{H}_{i}}}{2a_{i}^{2}}-{{\beta }_{i}}{{\sigma }_{i}}{{\hat{\theta }}_{i}} \label{API}
\end{align}
where ${{\sigma }_{i}}>0$ is designed constant. It's easy to see that ${{k}_{zi}}{{g}_{i{{\iota }_{i}}}}{{\alpha }_{i}}\le {{k}_{zi}}{{g}_{i0}}{{\alpha }_{i}}$. Substituting (\ref{IEI1})-(\ref{IEI2}), (\ref{VCi}) and (\ref{API}) into (\ref{dNLFi}) yields
\begin{align}
{{{\dot{V}}}_{i}}\le& {{{\dot{V}}}_{i-1}}+\frac{{{g}_{\left( i+1 \right)0}}k_{zi}^{2}z_{i+1}^{2}}{2}+\frac{{{g}_{i0}}\theta _{i}^{*}{{H}_{i}}{{k}_{zi}}}{2a_{i}^{2}}+{{g}_{i0}}{{k}_{zi}}{{\phi }_{i}}+{{k}_{zi}}{{g}_{i0}}{{\alpha }_{i}}+\frac{g_{i1}^{2}}{2{{g}_{\left( i+1 \right)0}}}+\frac{a_{i}^{2}}{2{{g}_{i0}}} \nonumber \\
&+\frac{1}{2{{g}_{i0}}}\sum\limits_{j=1}^{i}{\left( \varepsilon {{_{j}^{*}}^{2}}+D_{j}^{2} \right)}+\frac{1}{2{{g}_{i0}}}\sum\limits_{j=1}^{i-1}{g_{j1}^{2}}+\frac{4i-1}{2{{g}_{i0}}}-{{g}_{i0}}{{{\tilde{\theta }}}_{i}}\left( \frac{{{k}_{zi}}{{H}_{i}}}{2a_{i}^{2}}-{{\sigma }_{i}}{{{\hat{\theta }}}_{i}} \right) \nonumber \\
=&{{{\dot{V}}}_{i-1}}+\frac{{{g}_{\left( i+1 \right)0}}k_{zi}^{2}z_{i+1}^{2}}{2}-{{K}_{i}}{{k}_{zi}}{{z}_{i}}{{g}_{i0}}-\frac{{{g}_{i0}}k_{z\left( i-1 \right)}^{2}z_{i}^{2}}{2}-{{g}_{i0}}{{\sigma }_{i}}\frac{\tilde{\theta }_{i}^{2}}{2}+{{\Gamma }_{i}} \label{dNLFi2}
\end{align}
where ${{\Gamma }_{i}}=\frac{g_{i1}^{2}}{2{{g}_{\left( i+1 \right)0}}}+\frac{a_{i}^{2}}{2{{g}_{i0}}}+\frac{1}{2{{g}_{i0}}}\sum\limits_{j=1}^{i}{\left( \varepsilon {{_{j}^{*}}^{2}}+D_{j}^{2} \right)}+\frac{1}{2{{g}_{i0}}}\sum\limits_{j=1}^{i-1}{g_{j1}^{2}}+\frac{4i-1}{2{{g}_{i0}}}+{{g}_{i0}}{{\sigma }_{i}}\frac{\theta _{i}^{*2}}{2}$. Since
\begin{align}
{{\dot{V}}_{i-1}}\le \frac{{{g}_{i0}}k_{z\left( i-1 \right)}^{2}z_{i}^{2}}{2}-{{K}_{1}}{{g}_{10}}e{{(t)}^{2}}-\sum\limits_{j=2}^{i-1}{{{K}_{j}}{{k}_{zj}}{{z}_{j}}{{g}_{j0}}}-\sum\limits_{j=1}^{i-1}{{{g}_{j0}}{{\sigma }_{j}}\frac{\tilde{\theta }_{j}^{2}}{2}}+\sum\limits_{j=1}^{i-1}{{{\Gamma }_{j}}} \label{dVII}
\end{align}
substituting (\ref{dVII}) into (\ref{dNLFi2}) yields
\begin{align}
{{\dot{V}}_{i}}\le \frac{{{g}_{\left( i+1 \right)0}}k_{zi}^{2}z_{i+1}^{2}}{2}-{{K}_{1}}{{g}_{10}}e{{(t)}^{2}}-\sum\limits_{j=2}^{i}{{{K}_{j}}{{k}_{zj}}{{z}_{j}}{{g}_{j0}}}-\sum\limits_{j=1}^{i}{{{g}_{j0}}{{\sigma }_{j}}\frac{\tilde{\theta }_{j}^{2}}{2}}+\sum\limits_{j=1}^{i}{{{\Gamma }_{j}}}
\end{align}

\textbf{Step $n$ :} The time derivative of $z_n$ is
\begin{align}
{{\dot{z}}_{n}}={{f}_{n}}({{\bar{x}}_{n}},0)+{{g}_{n{{\iota }_{n}}}}u-{{\dot{\alpha }}_{n-1}}+{{d}_{n}}(t)
\end{align}
where ${{z}_{n}}={{x}_{n}}-{{\alpha }_{n-1}}$ and $\alpha_{n-1}$ is virtual control. And ${{f}_{n}}({{\bar{x}}_{n}},0)=W_{n}^{T}{{S}_{n}}\left( {{Z}_{n}} \right)+{{\varepsilon }_{n}}$ by Lemma 1, where $W_{n}^{T}$ is the optimal weight vector and ${{\varepsilon }_{n}}$ is approximation error satisfying $\left| {{\varepsilon }_{n}} \right|\le {{\varepsilon }_{n}^{*}}$. Define the integral Barrier Lyapunov function as
\begin{align}
{{V}_{n}}={{V}_{n-1}}+\int_{0}^{{{z}_{n}}}{\frac{\sigma k_{cn}^{2}}{k_{cn}^{2}-{{\left( \sigma +{{\alpha }_{n-1}} \right)}^{2}}}}d\sigma +\frac{{{g}_{n0}}}{{{\beta }_{n}}}\tilde{\theta }_{n}^{2} \label{LFN}
\end{align}
where $\beta_n>0$ is defined constant, ${{\tilde{\theta }}_{n}}=\theta _{n}^{*}-{{\hat{\theta }}_{n}}$ and ${{\hat{\theta }}_{n}}$ is the approximation of $\theta _{n}^{*}$. Define $\theta _{n}^{*}$ as $\theta _{n}^{*}=\max \left\{ {{\left\| {{W}_{1}} \right\|}^{2}},{{\left\| {{W}_{2}} \right\|}^{2}},\ldots ,{{\left\| {{W}_{n}} \right\|}^{2}} \right\}$. Similar to the first $n-1$ steps, system input $u$ is designed as
\begin{align}
u=-{{K}_{n}}{{z}_{n}}-\frac{k_{z\left( n-1 \right)}^{2}{{z}_{n}}}{2}\left( \frac{k_{cn}^{2}-x_{n}^{2}}{k_{cn}^{2}} \right)-{{\phi }_{n}}-\frac{{{{\hat{\theta }}}_{n}}{{H }_{n}}}{2a_{n}^{2}} \label{SI}
\end{align}
where $a_n$ is positive designed constant, $\phi_n$ and $H_n$ are the functions of all signals of the closed-loop system, which can be expressed as
\begin{align}
{{\phi }_{n}}=&{{k}_{zn}}+\frac{{{z}_{n}}\rho _{n-1}^{2}}{2}\sum\limits_{j=1}^{n-1}{{{\left\| \frac{\partial {{\alpha }_{n-1}}}{\partial {{x}_{j}}}{{x}_{j+1}} \right\|}^{2}}\left( \frac{k_{cn}^{2}-x_{n}^{2}}{k_{cn}^{2}} \right)}+{{z}_{n}}\rho _{n-1}^{2}\sum\limits_{j=1}^{n-1}{{{\left( \frac{\partial {{\alpha }_{n-1}}}{\partial {{x}_{j}}} \right)}^{2}}}\left( \frac{k_{cn}^{2}-x_{n}^{2}}{k_{cn}^{2}} \right) \nonumber \\
&+\frac{{{z}_{n}}\rho _{n-1}^{2}}{2}\sum\limits_{j=1}^{n-1}{{{\left\| \frac{\partial {{\alpha }_{n-1}}}{\partial y_{d}^{\left( j \right)}}y_{d}^{\left( j+1 \right)} \right\|}^{2}}\left( \frac{k_{cn}^{2}-x_{n}^{2}}{k_{cn}^{2}} \right)+}\frac{{{z}_{n}}\rho _{n-1}^{2}}{2}\sum\limits_{j=1}^{n-1}{{{\left\| \frac{\partial {{\alpha }_{n-1}}}{\partial {{\mu }^{\left( j \right)}}}{{\mu }^{\left( j+1 \right)}} \right\|}^{2}}}\left( \frac{k_{cn}^{2}-x_{n}^{2}}{k_{cn}^{2}} \right) \nonumber \\
&+\frac{{{z}_{n}}\rho _{n-1}^{2}}{2}\sum\limits_{j=1}^{n-1}{{{\left\| \frac{\partial {{\alpha }_{n-1}}}{\partial {{{\hat{\theta }}}_{j}}}{{{\dot{\hat{\theta }}}}_{j}} \right\|}^{2}}}\left( \frac{k_{cn}^{2}-x_{n}^{2}}{k_{cn}^{2}} \right) \label{PHI}\\
{{H}_{n}}=&{{k}_{zn}}S_{n}^{T}{{S}_{n}}+{{z}_{n}}\rho _{n-1}^{2}\sum\limits_{j=1}^{n-1}{{{\left\| \frac{\partial {{\alpha }_{n-1}}}{\partial {{x}_{j}}}{{S}_{j}}\left( {{Z}_{j}} \right) \right\|}^{2}}}\left( \frac{k_{cn}^{2}-x_{n}^{2}}{k_{cn}^{2}} \right) \label{HN}
\end{align}
where ${{k}_{zn}}=\frac{k_{cn}^{2}{{z}_{n}}}{k_{cn}^{2}-x_{n}^{2}}$, ${{\rho }_{n-1}}=\frac{{{k}_{cn}}}{2{{z}_{n}}}\ln \frac{\left( {{k}_{cn}}+{{z}_{n}}+{{\alpha }_{n-1}} \right)\left( {{k}_{cn}}-{{\alpha }_{n-1}} \right)}{\left( {{k}_{cn}}-{{z}_{n}}-{{\alpha }_{n-1}} \right)\left( {{k}_{cn}}+{{\alpha }_{n-1}} \right)}$, $\rho_{n-1}$ is well-defined in the neighbor of $z_{n}=0$ when $\left| {{\alpha }_{n-1}} \right|\le {{k}_{cn}}$.

Design the adaptation parameter as
\begin{align}
{{\dot{\hat{\theta }}}_{n}}=\frac{{{\beta }_{n}}{{k}_{zn}}{{H}_{n}}}{2a_{n}^{2}}-{{\beta }_{n}}{{\sigma }_{n}}{{\hat{\theta }}_{n}},\label{APN}
\end{align}
similar to the construction and analysis process of the first $n-1$ steps, substituting (\ref{SI}) and (\ref{APN}) into the time derivative of (\ref{LFN}) yields
\begin{align}
{{\dot{V}}_{n}}\le -{{K}_{1}}{{g}_{10}}e{{(t)}^{2}}-\sum\limits_{j=2}^{n}{{{K}_{j}}{{k}_{zj}}{{z}_{j}}{{g}_{j0}}}-\sum\limits_{j=1}^{n}{{{g}_{j0}}{{\sigma }_{j}}\frac{\tilde{\theta }_{j}^{2}}{2}}+\sum\limits_{j=1}^{n}{{{\Gamma }_{j}}}, \label{FdLF}
\end{align}
where ${{\Gamma }_{n}}=\frac{a_{n}^{2}}{2{{g}_{n0}}}+\frac{1}{2{{g}_{n0}}}\sum\limits_{j=1}^{n}{\left( \varepsilon {{_{j}^{*}}^{2}}+D_{j}^{2} \right)}+\frac{1}{2{{g}_{n0}}}\sum\limits_{j=1}^{n-1}{g_{j1}^{2}}+\frac{4n-1}{2{{g}_{n0}}}+{{g}_{n0}}{{\sigma }_{n}}\frac{\theta _{n}^{*2}}{2}$. Since $\int_{0}^{{{z}_{i}}}{\frac{\sigma k_{ci}^{2}}{k_{ci}^{2}-{{\left( \sigma +{{\alpha }_{i-1}} \right)}^{2}}}}d\sigma \le \frac{k_{ci}^{2}z_{i}^{2}}{k_{ci}^{2}-x_{i}^{2}},i=2,\ldots ,n$ in the interval $|{\left( \sigma +{{\alpha }_{i-1}} \right)}|<k_{ci}$, (\ref{FdLF}) can be rewritten as
\begin{align}
{{\dot{V}}_{n}}\le -{{K}_{1}}{{g}_{10}}e{{(t)}^{2}}-\sum\limits_{j=2}^{n}{{{K}_{j}}{{g}_{j0}}{{V}_{zj}}}-\sum\limits_{j=1}^{n}{{{g}_{j0}}{{\sigma }_{j}}\frac{\tilde{\theta }_{j}^{2}}{2}}+\sum\limits_{j=1}^{n}{{{\Gamma }_{j}}}. \label{FFdLF}
\end{align}

Define $C=\min \{2{{K}_{1}}{{g}_{10}},{{K}_{i+1}}{{g}_{\left( i+1 \right)0}},2{{\sigma }_{i}}{{\beta }_{i}},2{{\sigma }_{n}}{{\beta }_{n}},i=1,\ldots ,n-1\}$, (\ref{FFdLF}) can be expressed as
\begin{align}
{{\dot{V}}_{n}}\le -C{{V}_{n}}+D,\label{FF}
\end{align}
where $D=\sum\limits_{j=1}^{n}{{{\Gamma }_{j}}}$. Integrating (\ref{FF}) yields
\begin{align}
{{V}_{n}}\left( t \right)\le \left( V(0)-\frac{D}{C} \right){{e}^{-Ct}}+\frac{D}{C}\le V(0){{e}^{-Ct}}+\frac{D}{C} \label{IV1}
\end{align}
thus, it's obvious that all signals of the closed-loop system are semi-global ultimately uniformly bounded. When $\left| {{x}_{i}}\left( 0 \right) \right|<{{k}_{ci}},\left| {{\alpha }_{i-1}} \right|<{{k}_{ci}},i=2,\ldots ,n$, $V_0$ is bounded. since $V_n(t)$ is bounded, $\forall t>0,\left| {{x}_{i}} \right|<{{k}_{ci}},i=2,\ldots ,n$. Define $A_0$ is the bound of desired output signal $y_d$, select appropriate parameters of $\mu(t)$ to guarantee ${{A}_{0}}+\mu \left( 0 \right)<{{k}_{c1}}$, which makes sure $\forall t>0,\left| {{x}_{1}} \right|<{{k}_{c1}}$. Therefore, the whole state variables remain within the predefined constraints.

In real pure-feedback nonlinear systems, the sign of ${\partial f({{{\bar{x}}}_{n}},u)}/{\partial u}\;$ is unknown. To solve this problem, we relax Assumption 1, i.e., $0<{{g}_{n0}}<\left| {{g}_{n}} \right|<{{g}_{n1}}$. Inspired by Lemma 4, we redesign the system input with unknown control direction as
\begin{align}
u&=N\left( \zeta  \right)\left( {{K}_{n}}{{z}_{n}}+\frac{k_{z\left( n-1 \right)}^{2}{{z}_{n}}}{2}\left( \frac{k_{cn}^{2}-x_{n}^{2}}{k_{cn}^{2}} \right)+{{\phi }_{n}}\text{+}\frac{{{{\hat{\theta }}}_{n}}{{H }_{n}}}{2a_{n}^{2}} \right) \label{NNU}\\
\dot{\zeta }&={{K}_{n}}{{k}_{zn}}{{z}_{n}}+\frac{k_{z\left( n-1 \right)}^{2}k_{zn}z_{n}}{2}+\frac{{{k}_{zn}}{{{\hat{\theta }}}_{n}}{{H }_{n}}}{2a_{n}^{2}}+{{k}_{zn}}{{\phi }_{n}} \label{NUSS}
\end{align}
where $N\left( \zeta  \right)$ is Nussbaum-type even function, i.e., $N\left( \zeta  \right)\text{=}{{e}^{{{\zeta }^{2}}}}\cos \left( \left( {\pi }/{2}\; \right)\zeta  \right)$, and ${{\phi }_{n}},{{H}_{n}},{{\hat{\theta }}_{n}}$ have the same expression as (\ref{PHI})-(\ref{APN}) do. Redesign $V_n$ as
\begin{align}
{{V}_{n}}={{V}_{n-1}}+\int_{0}^{{{z}_{n}}}{\frac{\sigma k_{cn}^{2}}{k_{cn}^{2}-{{\left( \sigma +{{\alpha }_{n-1}} \right)}^{2}}}}d\sigma +\frac{1}{{{\beta }_{n}}}\tilde{\theta }_{n}^{2} \label{NLFN}
\end{align}
similar to the $n$th step, after inequality scaling, substituting (\ref{NNU}) and (\ref{NUSS}) into the time derivative of (\ref{NLFN}) yields
\begin{align}
{{{\dot{V}}}_{n}}\le&-{{K}_{1}}{{g}_{10}}e{{(t)}^{2}}-\sum\limits_{j=2}^{n-1}{{{K}_{j}}{{g}_{j0}}{{V}_{zj}}}+\left( N\left( \zeta  \right){{g}_{n{{\iota }_{n}}}}+1 \right)\dot{\zeta }-{{K}_{n}}{{V}_{zn}}  \nonumber \\
&-\sum\limits_{j=1}^{n-1}{{{g}_{j0}}{{\sigma }_{j}}\frac{\tilde{\theta }_{j}^{2}}{2}}-{{\sigma }_{n}}\frac{\tilde{\theta }_{n}^{2}}{2}+\sum\limits_{j=1}^{n}{{{\Gamma }_{j}}} \label{dNLFN}
\end{align}
Define $\eta =\min \{2{{K}_{1}}{{g}_{10}},{{K}_{i+1}}{{g}_{\left( i+1 \right)0}},2{{\sigma }_{j}}{{\beta }_{j}},{{K}_{n}},i=1,\ldots ,n-2,j=1,\ldots ,n\}$, (\ref{dNLFN}) can be expressed as
\begin{align}
{{\dot{V}}_{n}}\le \left( N\left( \zeta  \right){{g}_{n{{\iota }_{n}}}}+1 \right)\dot{\zeta }-\eta {{V}_{n}}+\rho,\label{FdNLFN}
\end{align}
where $\rho=\sum\limits_{j=1}^{n}{{{\Gamma }_{j}}}$. Integrating (\ref{FdNLFN}) yields
\begin{align}
{{V}_{n}}\left( t \right)\le {{V}_{n}}\left( 0 \right)+\int_{0}^{t}{{{e}^{\eta \left( \tau -t \right)}}\left( N\left( \zeta  \right){{g}_{n{{\iota }_{n}}}}+1 \right)\dot{\zeta }}d\tau +\frac{\rho }{\eta }. \label{IV2}
\end{align}
With the aid of Lemma 4, $V_n(t)$ and $\zeta(t)$ are bounded.

All in all, if $\left| {{x}_{i}}\left( 0 \right) \right|<{{k}_{ci}},\left| {{\alpha }_{i-1}} \right|<{{k}_{ci}},i=2,\ldots ,n$, $V_0$ is bounded. Thus from (\ref{IV1}) and (\ref{IV2}), $V_n(t)$ is bounded, $\forall t>0,\left| {{x}_{i}} \right|<{{k}_{ci}},i=2,\ldots ,n$. All the signals of the closed-loop system are semi-global ultimately uniformly bounded. Select appropriate parameters of $\mu(t)$ to guarantee ${{A}_{0}}+\mu \left( 0 \right)<{{k}_{c1}}$, which makes sure $\forall t>0,\left| {{x}_{1}} \right|<{{k}_{c1}}$. Therefore, the output tracking error converges to the preset arbitrarily small bound $\mu_{T_0}$ within the prescribed finite-time interval $T_0$ without overshooting predefined maximum, and the whole state variables remain within the preset constraints.
\section{Feasibility Check}
The above derivation and analysis process of integral Barrier Lyapunov functions assumes ${{k}_{ci}}>\left| {{\alpha }_{i-1}} \right|,i=2,\ldots ,n$ in the set ${{\Omega }}=\{{{\bar{z}}_{n}}\in {{\mathbb{R}}^{n}},{{\bar{y}}_{d}}\in {{\mathbb{R}}^{n+1}}:|{{z}_{i}}|\le \sqrt{2V(t)},|{{y}_{d}}|\le {{A}_{0}},|y_{d}^{(i)}|\le {{A}_{i}},i=1,\ldots ,n\}$. It's necessary to take feasibility check as a priori. Define a set of controller parameters to be optimized as $\kappa ={{\left[ {{K}_{1}},\ldots ,{{K}_{n-1}} \right]}^{T}}$, which are related to bounds of virtual controls and the convergent rate of the closed-loop system. Thus, we need to check if there exists a solution $\kappa ={{\left[ {{K}_{1}},\ldots ,{{K}_{n-1}} \right]}^{T}}$ for the following static semi-infinite nonlinear constrained problem
\begin{align}
\underset{{{K}_{1}},\ldots ,{{K}_{n-1}}>0}{\mathop{\max }}\,N\left( \kappa  \right)=\sum\limits_{j=1}^{n-1}{{{K}_{j}}}
\end{align}
subject to
\begin{align}
{{k}_{ci}}>\underset{({{\bar{z}}_{n}},{{\bar{y}}_{d}})\in {{\Omega }}}{\mathop{\sup }}\,\left| {{\alpha }_{i-1}}\left( \kappa  \right) \right|,i=2,\ldots ,n
\end{align}
\section{Simulation Illustration}
In this section, two numerical examples are provided as follows to demonstrate the effectiveness of proposed control method.

\textit{Example 1:} Considering the pure-feedback nonlinear system with full state constraints
\begin{equation}
{\left\{ \begin{array}{l}
{{{\dot{x}}}_{1}}=0.1{{x}_{1}}+{{x}_{2}}+d_{1}(t) \\
{{{\dot{x}}}_{2}}=0.1{{x}_{1}}{{x}_{2}}-0.2{{x}_{1}}+\left( 1+x_{1}^{2} \right)u\left( t \right)+d_{2}(t) \\
y={{x}_{1}}
\end{array} \right.}
\end{equation}
where $x_1,x_2$ are state variables, $u$ and $y$ are input and output of the system, respectively. $d_1(t)=0.5\cos \left( t \right),d_{2}=0.5\cos \left( 10t \right)$, and the desired output signal $y_d=2cos(t)$. The state variables are constrained by $\left|x_1\right|<3.2,\left|x_2\right|<8$.

Considering the order of the system is $2$, which means we should select $0<\lambda<{1}/{2}\;$ to avoid singularity of controllers, we select $\lambda=0.3$. To guarantee $\mu(0)+A_{0}<k_{c1}$, the other parameters of $\mu(t)$ are chosen as $\mu_{T_{0}}=0.05,\mu_{0}=1,\tau=1$. Thus ${{T}_{0}}={\mu _{0}^{\lambda }}/{\lambda \tau }\;=3.33s,\mu{0}+A_{0}=3.05<k_{c1}$ and $\underset{t\to 3.33s}{\mathop{\lim }}\,\mu \left( t \right)=0.05,\forall t>3.33s,\mu(t)=0.05$, which infers $\forall t>3.33s$, output tracking error $z_1\le 0.05$. The controllers and adaptation laws are given as follows
\begin{align}
{{\alpha }_{1}}&=-\frac{{{K}_{1}}e(t)}{\varphi }-\frac{{{{\hat{\theta }}}_{1}}e(t)\varphi S_{1}(Z_{1})^{T}{{S}_{1}(Z_{1})}}{2a_{1}^{2}}-\frac{e(t)\varphi {{{\left( {{{\dot{y}}}_{d}} \right)}^{2}}}}{2}-e(t)\varphi -\frac{e(t){{\Phi }^{2}}}{2\varphi } \\
u&=-{{K}_{2}}{{z}_{2}}-\frac{e{{(t)}^{2}}\varphi^{2}{{z}_{2}}}{2}\left( \frac{k_{c2}^{2}-x_{2}^{2}}{k_{c2}^{2}} \right)-{{\phi }_{2}}-\frac{{{{\hat{\theta }}}_{2}}{{H }_{2}}}{2a_{2}^{2}} \\
{{\dot{\hat{\theta }}}_{1}}&=\frac{{{\beta }_{1}}e{{(t)}^{2}}{{\varphi }^{2}}S_{1}(Z_{1})^{T}{{S}_{1}(Z_{1})}}{2a_{1}^{2}}-{{\beta }_{1}}{{\sigma }_{1}}{{\hat{\theta }}_{1}} \\
{{\dot{\hat{\theta }}}_{2}}&=\frac{{{\beta }_{2}}{{k}_{z2}}{{H}_{2}}}{2a_{2}^{2}}-{{\beta }_{2}}{{\sigma }_{2}}{{\hat{\theta }}_{2}}
\end{align}
where $\phi_2,H_2$ have the same expressions as (\ref{phi2}) and (\ref{H2}) do. ${{Z}_{1}}={{x}_{1}}\in \mathbb{R},{{Z}_{2}}={{[{{x}_{1}},{{x}_{2}}]}^{T}}\in {{\mathbb{R}}^{2}}$. With feasibility check, the parameters of the controllers can be chosen through optimization function fmincon.m in Matlab as $K_1=6.4,K_2=3.2,\beta_1=\beta_2=5,\sigma_1=\sigma_2=5$. The initial conditions are selected as $x_{1}(0)=2.5,x_{2}(0)=0.1,{\hat{\theta}}_1={\hat{\theta}}_2=0.2$.

\begin{figure}[htbp]
\begin{centering}
\includegraphics[scale=0.6]{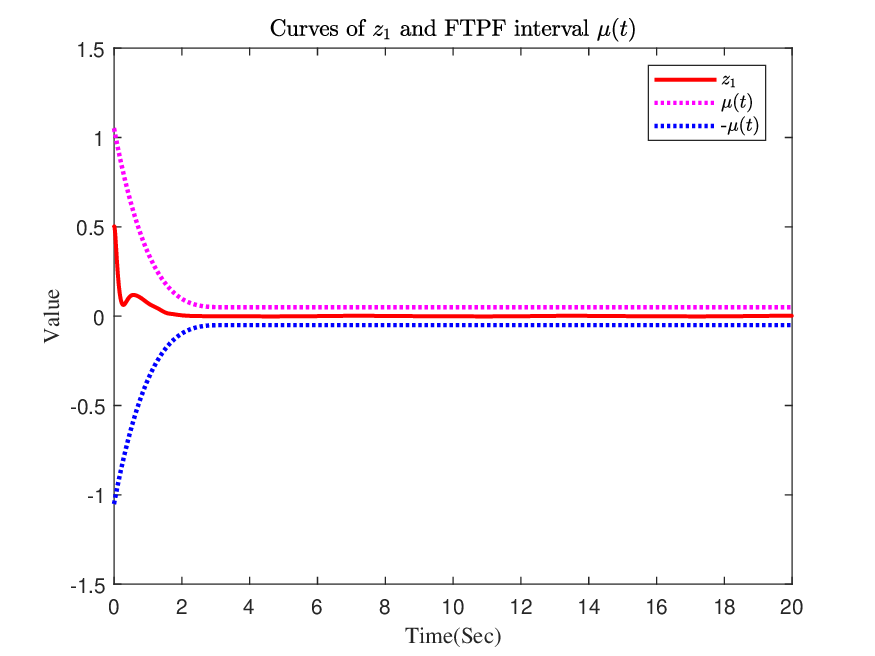}
\par\end{centering}
\caption{\label{P1}Curves of $z_1$ and interval of $\mu(t)$ and $-\mu(t)$.}
\end{figure}

\begin{figure}[htbp]
\begin{centering}
\includegraphics[scale=0.6]{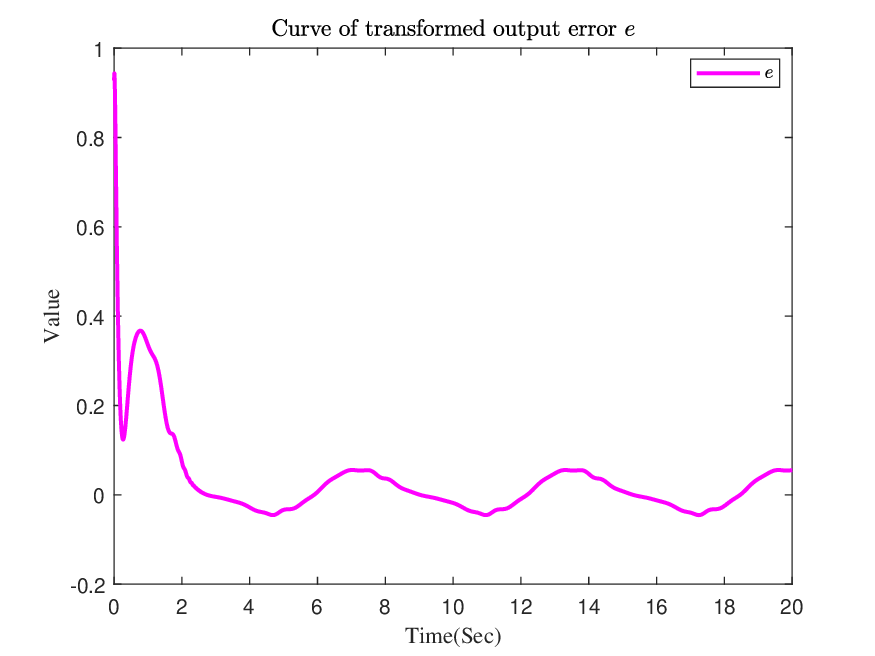}
\par\end{centering}
\caption{\label{P2}Curve of transformed output tracking error $e$.}
\end{figure}

\begin{figure}[htbp]
\begin{centering}
\includegraphics[scale=0.6]{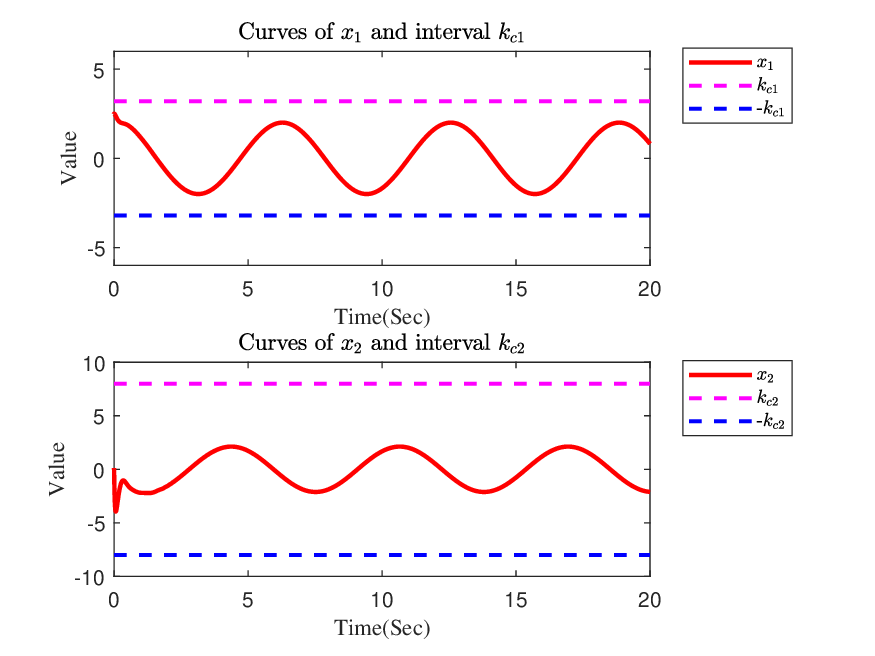}
\par\end{centering}
\caption{\label{P3}Curves of states $x_1,x_2$ and intervals $k_{c1},k_{c2}$.}
\end{figure}

\begin{figure}[htbp]
\begin{centering}
\includegraphics[scale=0.6]{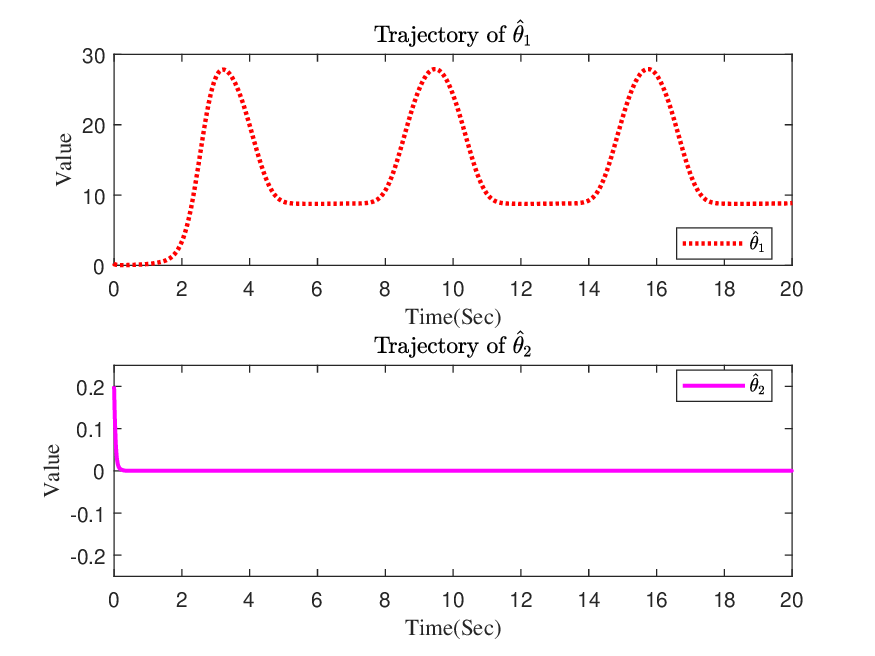}
\par\end{centering}
\caption{\label{P4}Curves of adaptation parameters ${\hat{\theta}}_1,{\hat{\theta}}_2$}
\end{figure}

\begin{figure}[htbp]
\begin{centering}
\includegraphics[scale=0.6]{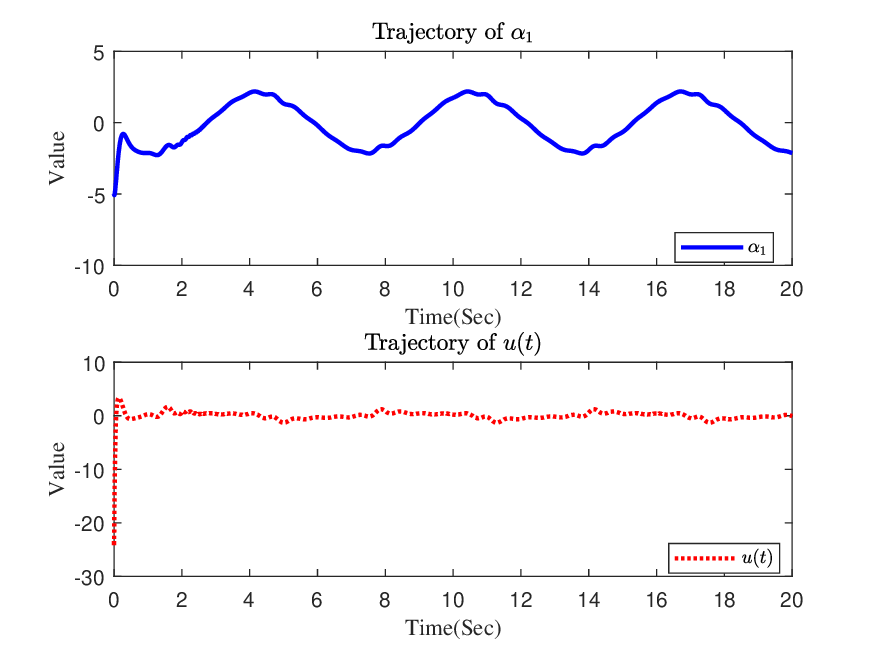}
\par\end{centering}
\caption{\label{P5}Curves of virtual control $\alpha_1$ and system input $u(t)$}
\end{figure}

The simulation results are shown in Figs. \ref{P1}-\ref{P5}. Fig. \ref{P1} depicts the curves of output tracking error $z_1$, which converges to predefined set in finite-time interval. Fig. \ref{P2} shows the trajectory of transformed output tracking error $e$. The state variables $x_{1},x_{2}$ are bounded in the predefined intervals $k_{c1}$ and $k_{c2}$ respectively in Fig. \ref{P3}. Fig. \ref{P4} shows the curves of adaptation parameters of two subsystems. Fig. \ref{P5} shows the trajectories of virtual control $\alpha_1$ and system input $u$.

\textit{Example 2:} Considering the inverted pendulum system with full state constraints
\begin{equation}
{\left\{ \begin{array}{l}
\dot{x}_1=x_2+d_1(t) \\
\dot{x}_2=\frac{g\sin(x_1)-\frac{mlx_2^2\cos(x_1)\sin(x_1)}{m+m_c}}{l\Big(\frac{4}{3}-\frac{m\cos^2(x_1)}{m+m_c}\Big)}+\frac{\frac{\cos(x_1)}{m+m_c}}{l\Big(\frac{4}{3}-\frac{m\cos^2(x_1)}{m+m_c}\Big)}u+d_2(t)\\
y=x_1
\end{array} \right.}
\end{equation}
where $x_1,x_2$ are the angle of the pendulum and the angular velocity, respectively, $u$ and $y$ denote input and output of the system, respectively, gravity coefficient $g=9.8$m/s$^{2}$, $m=0.1$kg and $m_c=1$kg represent the mass of a pole and the mass of a cart, respectively, and $l=0.5$m stand for the half length of a pole. $y_d=\sin(t)$ denotes the desired output signal. $d_1(t)=0.05\cos(t),d_2(t)=0.05\cos(10t)$. The state variables $x_1,x_2$ are constrained by $\left|x_1\right|<1.2$rad, $\left|x_2\right|<3.5$rad/s. The order of the system is $2$, to avoid singularity of controllers, we select $\lambda=0.3$, the other parameters of $\mu(t)$ are chosen as $\mu_{T_{0}}=0.01,\mu_{0}=1,\tau=1$ to guarantee $\mu(0)+A_{0}<k_{c1}$. Thus ${{T}_{0}}={\mu _{0}^{\lambda }}/{\lambda \tau }\;=3.33s,\mu{0}+A_{0}=3.01<k_{c1}$ and $\underset{t\to 3.33s}{\mathop{\lim }}\,\mu \left( t \right)=0.01,\forall t>3.33s,\mu(t)=0.01$, which infers $\forall t>3.33s$, output tracking error $z_1\le 0.01$rad. Since the sign of $g_2$, i.e., ${{f}_{2}}({{\bar{x}}_{2}},u)$ is unknown, the controllers and adaptation laws are given as follows
\begin{align}
{{\alpha }_{1}}&=-\frac{{{K}_{1}}e(t)}{\varphi }-\frac{{{{\hat{\theta }}}_{1}}e(t)\varphi S_{1}(Z_{1})^{T}{{S}_{1}(Z_{1})}}{2a_{1}^{2}}-\frac{e(t)\varphi {{{\left( {{{\dot{y}}}_{d}} \right)}^{2}}}}{2}-e(t)\varphi -\frac{e(t){{\Phi }^{2}}}{2\varphi } \\
u&=N\left( \zeta  \right)\left({{K}_{2}}{{z}_{2}}+\frac{e{{(t)}^{2}}\varphi^{2}{{z}_{2}}}{2}\left( \frac{k_{c2}^{2}-x_{2}^{2}}{k_{c2}^{2}} \right)+{{\phi }_{2}}+\frac{{{{\hat{\theta }}}_{2}}{{H }_{2}}}{2a_{2}^{2}}\right) \\
\dot{\zeta }&={{K}_{2}}k_{z2}{{z}_{2}}+\frac{e{{(t)}^{2}}\varphi^{2}k_{z2}{{z}_{2}}}{2}\left( \frac{k_{c2}^{2}-x_{2}^{2}}{k_{c2}^{2}} \right)+k_{z2}{{\phi }_{2}}+\frac{k_{z2}{{{\hat{\theta }}}_{2}}{{H }_{2}}}{2a_{2}^{2}} \\
{{\dot{\hat{\theta }}}_{1}}&=\frac{{{\beta }_{1}}e{{(t)}^{2}}{{\varphi }^{2}}S_{1}(Z_{1})^{T}{{S}_{1}(Z_{1})}}{2a_{1}^{2}}-{{\beta }_{1}}{{\sigma }_{1}}{{\hat{\theta }}_{1}} \\
{{\dot{\hat{\theta }}}_{2}}&=\frac{{{\beta }_{2}}{{k}_{z2}}{{H}_{2}}}{2a_{2}^{2}}-{{\beta }_{2}}{{\sigma }_{2}}{{\hat{\theta }}_{2}}
\end{align}
where $N\left( \zeta  \right)={{e}^{{{\zeta }^{2}}}}\cos ({\pi }/{2}\;\zeta )$,$\phi_2,H_2$ have the same expressions as (\ref{phi2}) and (\ref{H2}) do. ${{Z}_{1}}={{x}_{1}}\in \mathbb{R},{{Z}_{2}}={{[{{x}_{1}},{{x}_{2}}]}^{T}}\in {{\mathbb{R}}^{2}}$. With feasibility check, the parameters of the controllers can be chosen through optimization function fmincon.m in Matlab as $K_1=5.8,K_2=10,\beta_1=\beta_2=5,\sigma_1=\sigma_2=5$. The initial conditions are selected as $x_{1}(0)=0.01$rad,$x_{2}(0)=0.1$rad/s,${\hat{\theta}}_1={\hat{\theta}}_2=0.2,\zeta(0)=0.8$.

\begin{figure}[htbp]
\begin{centering}
\includegraphics[scale=0.6]{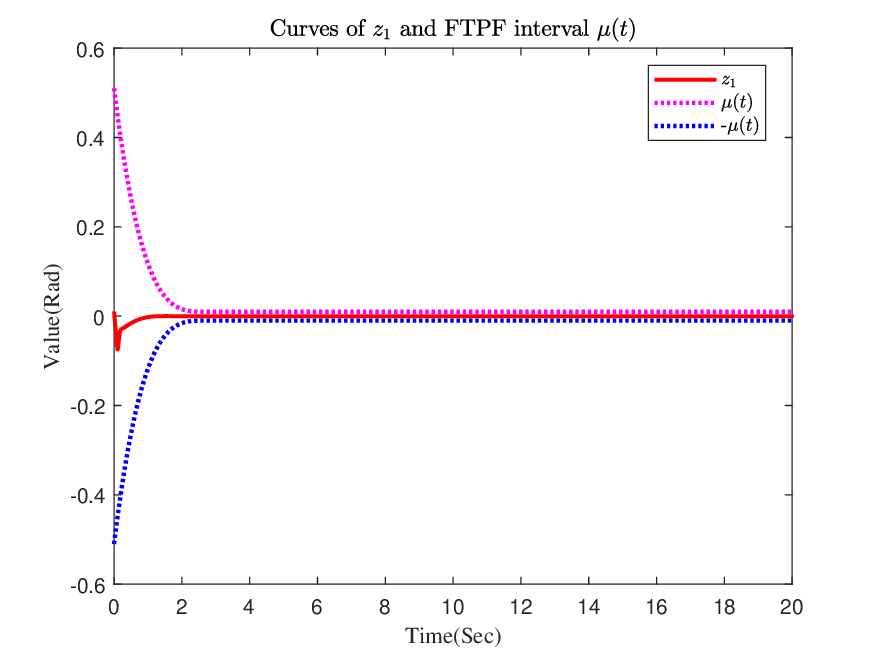}
\par\end{centering}
\caption{\label{P6}Curves of $z_1$ and interval of $\mu(t)$ and $-\mu(t)$.}
\end{figure}

\begin{figure}[htbp]
\begin{centering}
\includegraphics[scale=0.6]{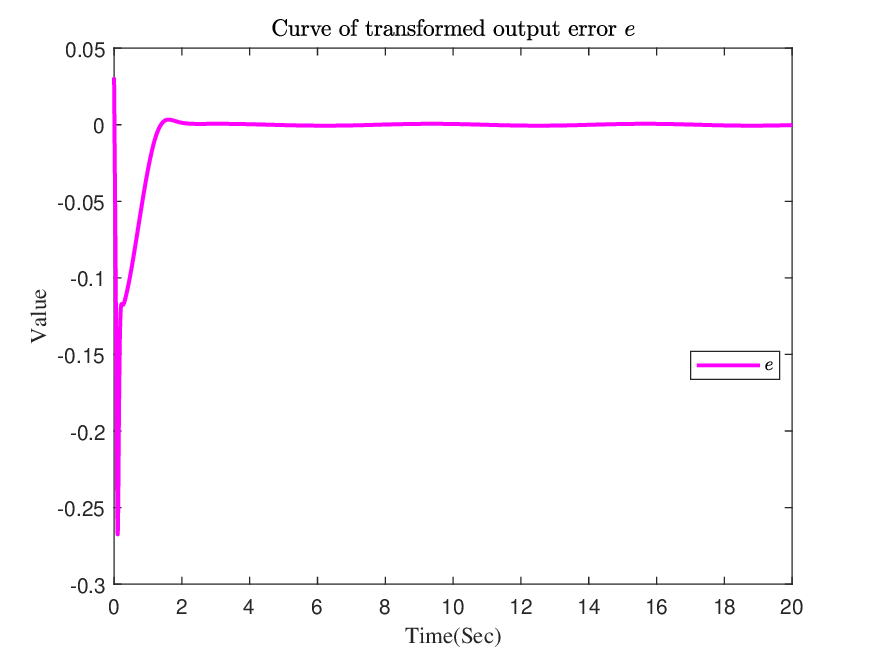}
\par\end{centering}
\caption{\label{P7}Curve of transformed output tracking error $e$.}
\end{figure}

\begin{figure}[htbp]
\begin{centering}
\includegraphics[scale=0.6]{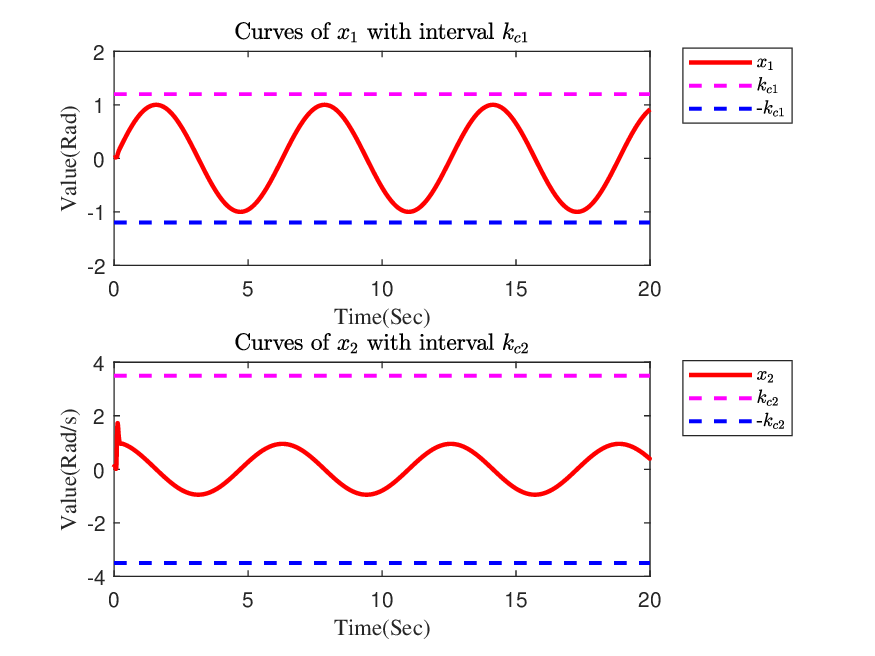}
\par\end{centering}
\caption{\label{P8}Curves of states $x_1,x_2$ and intervals $k_{c1},k_{c2}$.}
\end{figure}

\begin{figure}[htbp]
\begin{centering}
\includegraphics[scale=0.6]{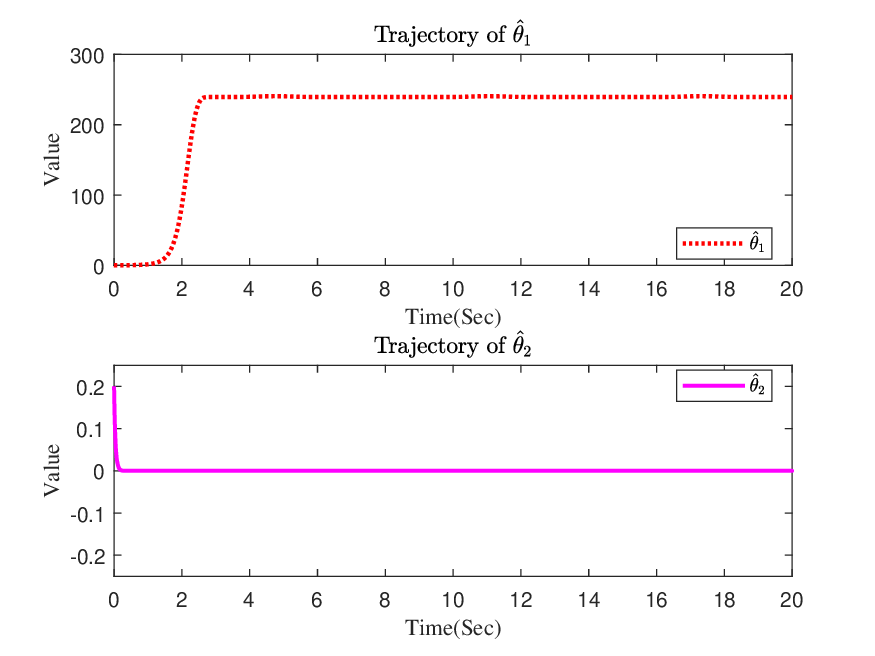}
\par\end{centering}
\caption{\label{P9}Curves of adaptation parameters ${\hat{\theta}}_1,{\hat{\theta}}_2$.}
\end{figure}

\begin{figure}[htbp]
\begin{centering}
\includegraphics[scale=0.6]{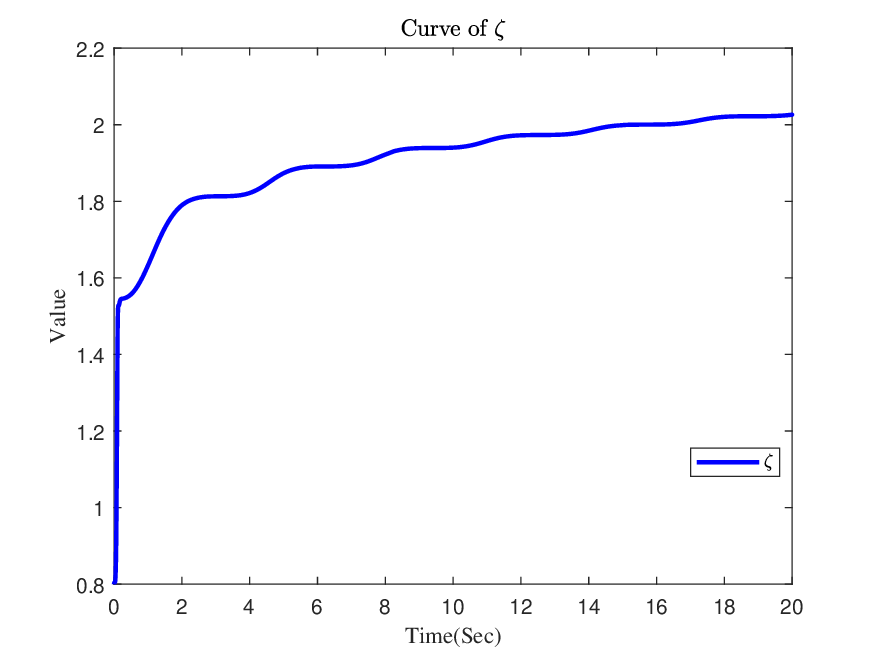}
\par\end{centering}
\caption{\label{P10}Curve of $\zeta(t)$.}
\end{figure}

\begin{figure}[htbp]
\begin{centering}
\includegraphics[scale=0.6]{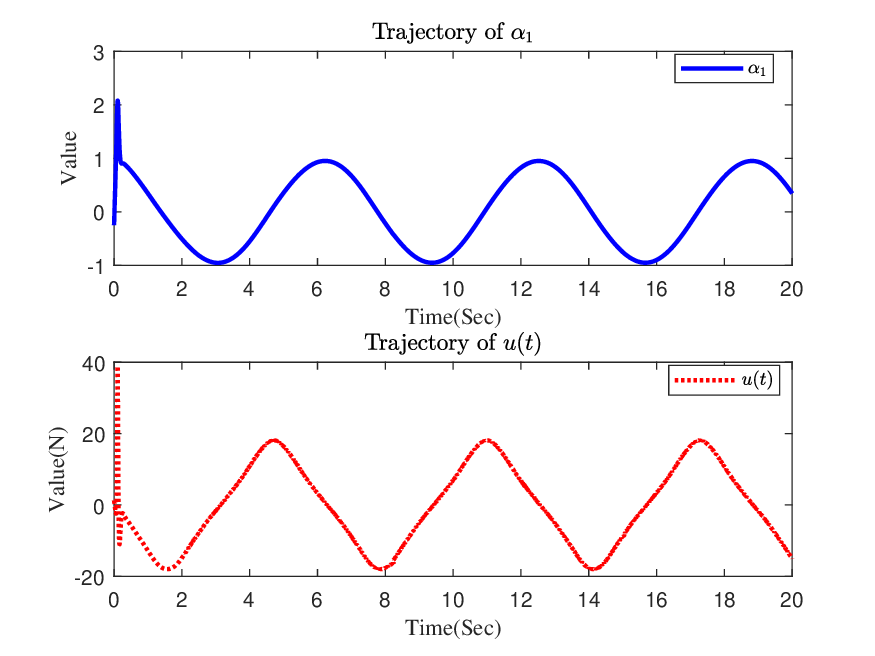}
\par\end{centering}
\caption{\label{P11}Curves of virtual control $\alpha_1$ and system input $u(t)$.}
\end{figure}

The simulation results are shown in Figs. \ref{P6}-\ref{P11}. Fig. \ref{P6} depicts the curves of output tracking error $z_1$, which converges to predefined set in finite-time interval. Fig. \ref{P7} shows the trajectory of transformed output tracking error $e$. The state variables $x_{1},x_{2}$ are bounded in the predefined intervals $k_{c1}$ and $k_{c2}$ respectively in Fig. \ref{P8}. Fig. \ref{P9} shows the curves of adaptation parameters of two subsystems. Fig. \ref{P10} shows the curve of $\zeta$. Fig. \ref{P11} shows the trajectories of virtual control $\alpha_1$ and system input $u$.

\section{Conclusion}
This paper studies the finite-time adaptive fuzzy tracking control problem for a class of pure-feedback nonlinear systems with full state constraint. The fuzzy logic systems are utilized to approximate unknown smooth functions. Carefully designed finite-time-stable like function is constructed to guarantee the output tracking error converges to the predefined set in the arbitrary finite interval. Integral Barrier Lyapunov functions are employed to deal with state constraints. Considering the sign of system input may be unknown, we redesign the system input with aid of Nussbaum-type function. By stability analysis, all the signals of the closed-loop system are semi-global ultimately uniformly bounded. Two simulation illustrations are performed to verify effectiveness of the developed method.




\end{document}